\documentclass{emulateapj}
\usepackage{lscape}

\newcommand{\EMint}{\ensuremath{\int \Ne^2 \dl}}

\newcommand{\Healpha}{He$\alpha$}

\newcommand{\Kalpha}{K$\alpha$}

\newcommand{\Lyalpha}{Ly$\alpha$}

\newcommand{\Ne}{\ensuremath{n_{\mathrm{e}}}}

\newcommand{\nH}{\ensuremath{n_{\mathrm{H}}}}

\newcommand{\nO}{\ensuremath{n_{\mathrm{O}}}}

\newcommand{\pLB}{\ensuremath{p_\mathrm{LB}}}
\newcommand{\rmin}{\ensuremath{r_\mathrm{min}}}

\newcommand{\RE}{\ensuremath{R_\mathrm{E}}}

\newcommand{\TLB}{\ensuremath{T_\mathrm{LB}}}
\newcommand{\logTLB}{\ensuremath{\log(\TLB/\mathrm{K})}}
\newcommand{\logTLBsq}{\ensuremath{\log[\TLB/\mathrm{K}]}}
\newcommand{\Thalo}{\ensuremath{T_\mathrm{halo}}}
\newcommand{\logThalo}{\ensuremath{\log(\Thalo/\mathrm{K})}}

\newcommand{\nm}{\ensuremath{\mbox{\nm}}}
\newcommand{\cm}{\ensuremath{\mbox{cm}}}

\newcommand{\km}{\ensuremath{\mbox{km}}}
\newcommand{\AU}{\ensuremath{\mbox{AU}}}
\newcommand{\pc}{\ensuremath{\mbox{pc}}}

\newcommand{\s}{\ensuremath{\mbox{s}}}

\newcommand{\kev}{\ensuremath{\mbox{keV}}}

\newcommand{\erg}{\ensuremath{\mbox{erg}}}

\newcommand{\sr}{\ensuremath{\mbox{sr}}}

\newcommand{\K}{\ensuremath{\mbox{K}}}

\newcommand{\ph}{\ensuremath{\mbox{photons}}}

\newcommand{\cmsq}{\ensuremath{\cm^2}}

\newcommand{\pcc}{\ensuremath{\cm^{-3}}}
\newcommand{\pcmsq}{\ensuremath{\cm^{-2}}}

\newcommand{\ps}{\ensuremath{\s^{-1}}}
\newcommand{\psr}{\ensuremath{\sr^{-1}}}

\newcommand{\emismeas}{\ensuremath{\cm^{-6}} \pc}

\newcommand{\flux}{\erg\ \pcmsq\ \ps}
\newcommand{\lineunit}{\ph\ \pcmsq\ \ps\ \psr}

\newcommand{\LU}{\ensuremath{\mbox{L.U.}}}
\newcommand{\kmps}{\km\ \ps}

\newcommand{\pownorm}{\ph\ \pcmsq\ \ps\ \psr\ \kev$^{-1}$}

\newcommand{\presalt}{\pcc\ \K}

\newcommand{\MgXI}{Mg~\textsc{xi}}

\newcommand{\NVII}{N~\textsc{vii}}

\newcommand{\NeIX}{Ne~\textsc{ix}}

\newcommand{\OVI}{O~\textsc{vi}}
\newcommand{\OVII}{O~\textsc{vii}}
\newcommand{\OVIII}{O~\textsc{viii}}

\newcommand{\ace}{\textit{ACE}}

\newcommand{\chandra}{\textit{Chandra}}

\newcommand{\fuse}{\textit{FUSE}}

\newcommand{\iras}{\textit{IRAS}}

\newcommand{\rosat}{\textit{ROSAT}}

\newcommand{\suzaku}{\textit{Suzaku}}

\newcommand{\wind}{\textit{WIND}}
\newcommand{\xmm}{\textit{XMM-Newton}}

\newcommand{\citepossessive}[1]{\citeauthor{#1}'s \citeyearpar{#1}}

\newcommand{\dl}{\ensuremath{\mathrm{d}l}}
\newcommand{\e}{\ensuremath{\mathrm{e}}}

\newcommand{\chisq}{\ensuremath{\chi^2}}

\newcommand{\Oplussix}{O$^\mathrm{+6}$}
\newcommand{\Oplusseven}{O$^\mathrm{+7}$}
\newcommand{\Opluseight}{O$^\mathrm{+8}$}
\newcommand{\nn}{\ensuremath{n_\mathrm{n}}}
\newcommand{\nX}{\ensuremath{n_\mathrm{X}}}
\newcommand{\nsw}{\ensuremath{n_\mathrm{sw}}}
\newcommand{\usw}{\ensuremath{u_\mathrm{sw}}}
\newcommand{\NC}{\ensuremath{N_\mathrm{C}}}

\shorttitle{OBSERVING CHARGE EXCHANGE WITH \textit{SUZAKU} AND \textit{XMM-NEWTON}}
\shortauthors{HENLEY AND SHELTON}

\begin{document}

\title{Comparing \textit{Suzaku} and \textit{XMM-Newton} Observations of the Soft X-ray Background:
Evidence for Solar Wind Charge Exchange Emission}
\author{David B. Henley and Robin L. Shelton}
\affil{Department of Physics and Astronomy, University of Georgia, Athens, GA 30602}
\email{dbh@physast.uga.edu}

\begin{abstract}
We present an analysis of a pair of \suzaku\ spectra of the soft X-ray background (SXRB), obtained from pointings on and off a nearby shadowing
filament in the southern Galactic hemisphere. Because of the different Galactic column densities in the two pointing directions, the
observed emission from the Galactic halo has a different shape in the two spectra. We make use of this difference when modeling the spectra
to separate the absorbed halo emission from the unabsorbed foreground emission from the Local Bubble (LB).
The temperatures and emission measures we obtain are significantly different from those determined
from an earlier analysis of \xmm\ spectra from the same pointing directions. We attribute this difference to the presence
of previously unrecognized solar wind charge exchange (SWCX) contamination in the \xmm\ spectra, possibly due to a localized enhancement
in the solar wind moving across the line of sight. Contemporaneous solar wind data from \ace\ show nothing unusual during the course
of the \xmm\ observations. Our results therefore suggest that simply examining contemporaneous solar wind data might be inadequate for
determining if a spectrum of the SXRB is contaminated by SWCX emission. If our \suzaku\ spectra are not
badly contaminated by SWCX emission, our best-fitting LB model gives a temperature of $\logTLB = 5.98^{+0.03}_{-0.04}$ and a pressure
of $\pLB / k = \mbox{13,100}$--16,100~\presalt. These values are lower than those obtained from other recent observations of the LB,
suggesting the LB may not be isothermal and may not be in pressure equilibrium. Our halo modeling, meanwhile, suggests that neon may
be enhanced relative to oxygen and iron, possibly because oxygen and iron are partly in dust.
\end{abstract}


\section{INTRODUCTION}
\label{sec:Introduction}

The diffuse soft X-ray background (SXRB), which is observed in all directions in the $\sim$0.1--2~\kev\ band, is composed of emission
from several different components. For many years, the observed 1/4-\kev\ emission was believed to originate from the Local Bubble (LB),
a cavity in the local interstellar medium (ISM) of $\sim$100~pc radius filled with $\sim$$10^6$~K gas \citep{sanders77,cox87,mccammon90,snowden90}.
However, the discovery of shadows in the 1/4-\kev\ background with \rosat\ showed that $\sim$50\%\ of the 1/4~\kev\ emission originates
from beyond the LB \citep{burrows91,snowden91}. This more distant emission originates from the Galactic halo, which contains hot gas
with temperatures $\logThalo \sim 6.0$--6.5 (\citealp{snowden98,kuntz00,smith07a,galeazzi07}; \citealp*{henley07}). As the halo gas is hotter than the LB
gas, it also emits at higher energies, up to $\sim$1~\kev. Above $\sim$1~\kev\ the X-ray background is extragalactic in origin, and
is due to unresolved active galactic nuclei (AGN; \citealp{mushotzky00}).

X-ray spectroscopy of the SXRB can, in principle, determine the thermal properties, ionization state, and chemical abundances of the
hot gas in the Galaxy. These properties give clues to the origin of the hot gas, which is currently uncertain. However, to determine the
physical properties of the hot Galactic gas, one must first decompose the SXRB into its constituents. This is achieved using a
technique called ``shadowing'', which makes use of the above-mentioned shadows cast in the SXRB by cool clouds of gas between the Earth
and the halo. Low-spectral-resolution \rosat\ observations of the SXRB were decomposed into their foreground and background components by
modeling the intensity variation due to the varying absorption column density on and around shadowing clouds
(\citealp{burrows91,snowden91,snowden00}; \citealp*{snowden93}; \citealp*{kuntz97}).
The same technique was used to decompose \rosat\ All-Sky Survey data over large areas of the sky \citep{snowden98,kuntz00}.

With higher resolution spectra, such as those from the CCD cameras onboard \xmm\ or \suzaku, it is possible to decompose the SXRB 
into its foreground and background components spectroscopically. This is achieved using one spectrum toward a shadowing cloud, and one
toward a pointing to the side of the cloud. The spectral shape of the absorbed background component (and hence of the overall spectrum) will differ
between the two directions, because of the different absorbing column densities. Therefore, by fitting a suitable multicomponent model
simultaneously to the two spectra, one can separate out the foreground and background emission components. Such a model will typically
consist of an unabsorbed single-temperature ($1T$) thermal plasma model for the LB, an absorbed thermal plasma model for the
Galactic halo, and an absorbed power-law for the extragalactic background. The Galactic halo model could be a $1T$ model, a two-temperature
($2T$) model, or a differential emission measure (DEM) model (\citealp{galeazzi07,henley07}; S.~J. Lei et al., in preparation).

Recent work has shown that there is an additional complication, as X-ray emission can originate within the solar system, via solar wind charge
exchange (SWCX; \citealp{cox98,cravens00}). In this process, highly ionized species in the solar wind interact with neutral atoms within the solar
system. An electron transfers from a neutral atom into an excited energy level of a solar wind ion, which then decays radiatively, emitting an
X-ray photon. The neutral atoms may be in the outer reaches of the Earth's atmosphere (giving rise to geocoronal emission), or they may be in
interstellar material flowing through the solar system (giving rise to heliospheric emission). It has been estimated that the heliospheric emission
may contribute up to $\sim$50\%\ of the observed soft X-ray flux \citep{cravens00}. The geocoronal emission is typically an order of magnitude
fainter, but during solar wind enhancements it can be of similar brightness to the heliospheric emission \citep{wargelin04}. SWCX line emission
has been observed with \chandra, \xmm, and \suzaku\ (\citealp{wargelin04}; \citealp*{snowden04}; \citealp{fujimoto07}). As the SWCX emission is time varying, it cannot easily
be modeled out of a spectrum of the SXRB.

\begin{figure}
\centering
\includegraphics[width=0.9\linewidth]{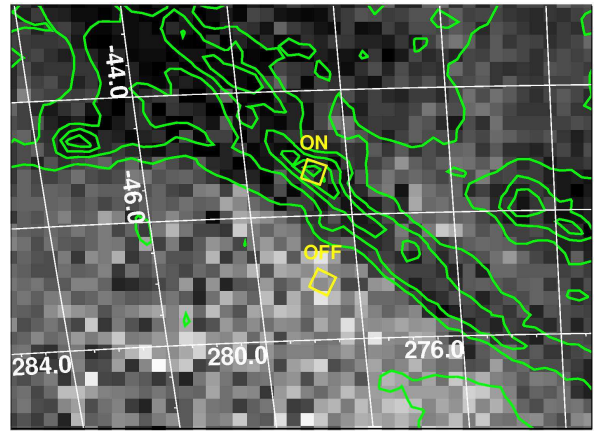}
\caption{The shadowing filament used for our observations, shown in Galactic coordinates.
\textit{Grayscale}: \rosat\ All-Sky Survey R1+R2 intensity \citep{snowden97}.
\textit{Contours}: \iras\ 100-micron intensity \citep{schlegel98}.
\textit{Yellow squares}: Our \suzaku\ pointing directions.
\label{fig:FilamentImage}}
\end{figure}

If SWCX contamination is not taken into account, analyses of SXRB spectra will yield incorrect results for the LB and halo gas.
This paper contains a demonstration of this fact.
\citet{henley07} analyzed a pair of \xmm\ spectra of the SXRB using the previously described shadowing technique. One spectrum was from a direction
toward a nearby shadowing filament in the southern Galactic hemisphere ($d = 230 \pm 30$~pc; \citealp{penprase98}), while the other was
from a direction $\sim$2\degr\ away. The filament and the shadow it casts in the 1/4-\kev\ background are shown in Figure~\ref{fig:FilamentImage}.
Contemporaneous solar wind data from the \textit{Advanced Composition Explorer} (\ace) showed that the solar wind was steady during
the \xmm\ observations, without any flares or spikes. The proton flux was slightly lower than average, and the oxygen ion ratios were fairly typical.
These observations led \citet{henley07} to conclude that
their spectra were unlikely to be severely contaminated by SWCX emission. Using a $2T$ halo model, they obtained a LB temperature of $\logTLB = 6.06$
and halo temperatures of $\logThalo = 5.93$ and 6.43. The LB temperature and the hotter halo temperature are in good agreement with other recent
measurements of the SXRB using \xmm\ and \suzaku\ \citep{galeazzi07,smith07a}, and with analysis of the \rosat\ All-Sky Survey \citep{kuntz00}.

We have obtained spectra of the SXRB from the same directions as \citepossessive{henley07} \xmm\ spectra with the X-ray Imaging Spectrometer (XIS;
\citealp{koyama07}) onboard the \suzaku\ X-ray observatory \citep{mitsuda07}. The XIS is an excellent tool for studying the SXRB, due to its low
non-X-ray background and good spectral resolution. Our \suzaku\ pointing
directions are shown in Figure~\ref{fig:FilamentImage}. We analyze our \suzaku\ spectra using the same shadowing technique used by
\citet{henley07}. We find that there is poor agreement between the results of our \suzaku\ analysis and the results
of the \xmm\ analysis in \citet{henley07}. We attribute this discrepancy to previously unrecognized SWCX contamination in the \xmm\ spectra,
which means that SWCX contamination can occur at times when the solar wind flux measured by \ace\ is low and does not show flares.

This paper is organized as follows. The \suzaku\ observations and data reduction are described in \S\ref{sec:DataReduction}.
The analysis of the spectra using multicomponent spectral models is described in \S\ref{sec:SuzakuAnalysis}. The discrepancy
between the \suzaku\ results and the \xmm\ results is discussed in \S\ref{sec:XMMAnalysis}. This discrepancy is due
to the presence of an additional emission component in the \xmm\ spectra, which we also describe in \S\ref{sec:XMMAnalysis}.
In \S\ref{sec:OxygenLines} we measure the total intensities of the oxygen lines in our \suzaku\ and \xmm\ spectra.
We concentrate on these lines because they are the brightest in our spectra, and are a major component of the 3/4-\kev\ SXRB \citep{mccammon02}.
In \S\ref{sec:SWCX} we present a simple model for estimating the intensity of the oxygen lines due to SWCX, which we
compare with our observations. We discuss our results in \S\ref{sec:Discussion}, and conclude with a summary in \S\ref{sec:Summary}.
Throughout this paper we quote $1\sigma$ errors.

\section{OBSERVATIONS AND DATA REDUCTION}
\label{sec:DataReduction}

Both of our observations were carried out in early 2006 March. The details of the observations are shown in
Table~\ref{tab:Observations}. In the following, we just analyze data from the back-illuminated XIS1 chip,
as it is more sensitive at lower energies than the three front-illuminated chips.

\begin{deluxetable*}{lcccccc}
\tablecaption{Details of Our \suzaku\ Observations\label{tab:Observations}}
\tablehead{
		& \colhead{Observation}	& \colhead{$l$}		& \colhead{$b$}		& \colhead{Start time}	& \colhead{End time}	& \colhead{Usable exposure} 	\\
		& \colhead{ID}		& \colhead{(deg)}	& \colhead{(deg)}	& \colhead{(UT)}	& \colhead{(UT)}	& \colhead{(ks)}
}
\startdata
Off filament	& 501001010		& 278.71		& $-47.07$		& 2006-03-01 16:56:01	& 2006-03-02 22:29:14	& 55.6				\\
On filament	& 501002010		& 278.65		& $-45.30$		& 2006-03-03 20:52:00	& 2006-03-06 08:01:19	& 69.0				\\
\enddata
\end{deluxetable*}

\begin{figure*}
\includegraphics[width=0.48\linewidth]{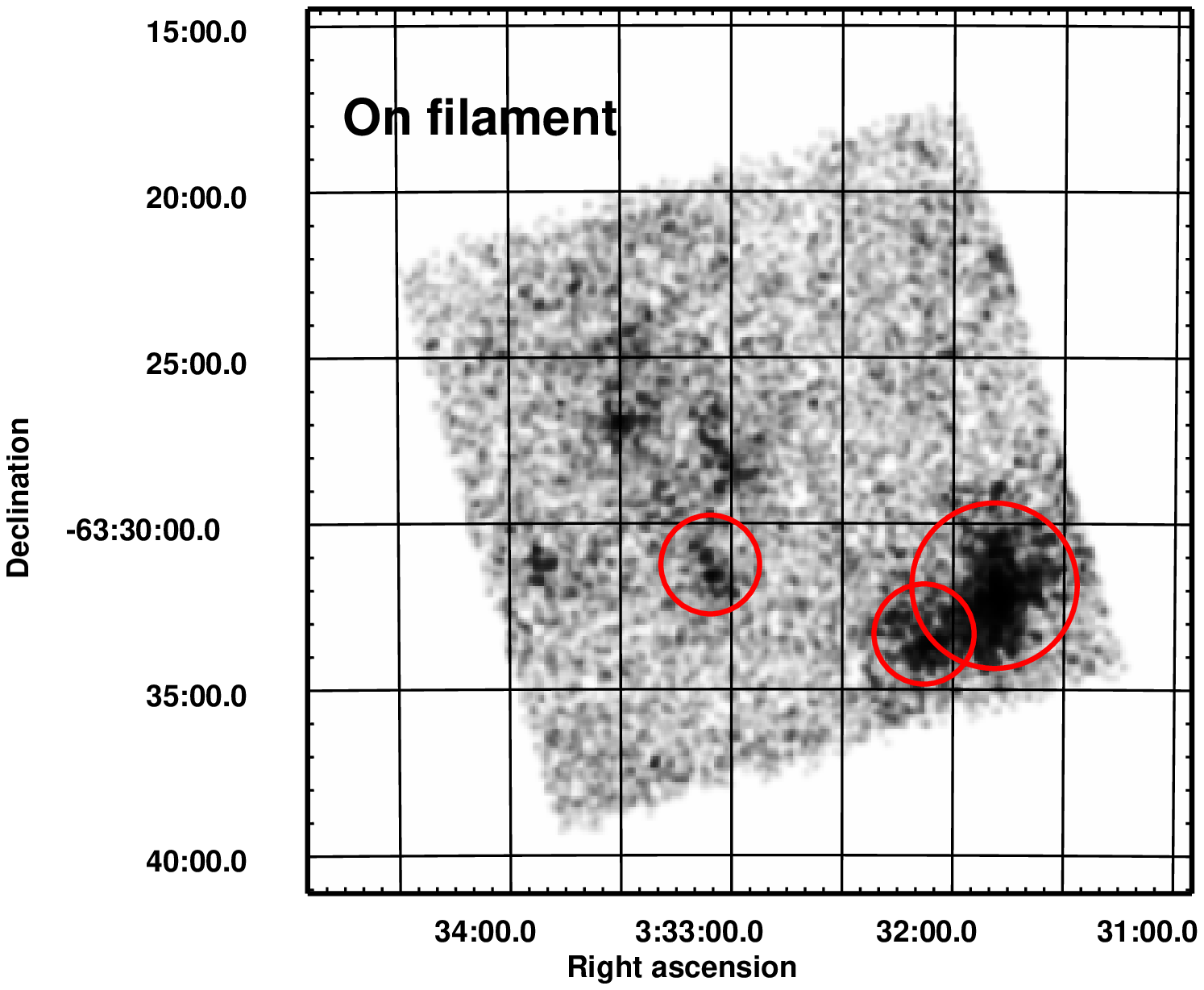}
\includegraphics[width=0.48\linewidth]{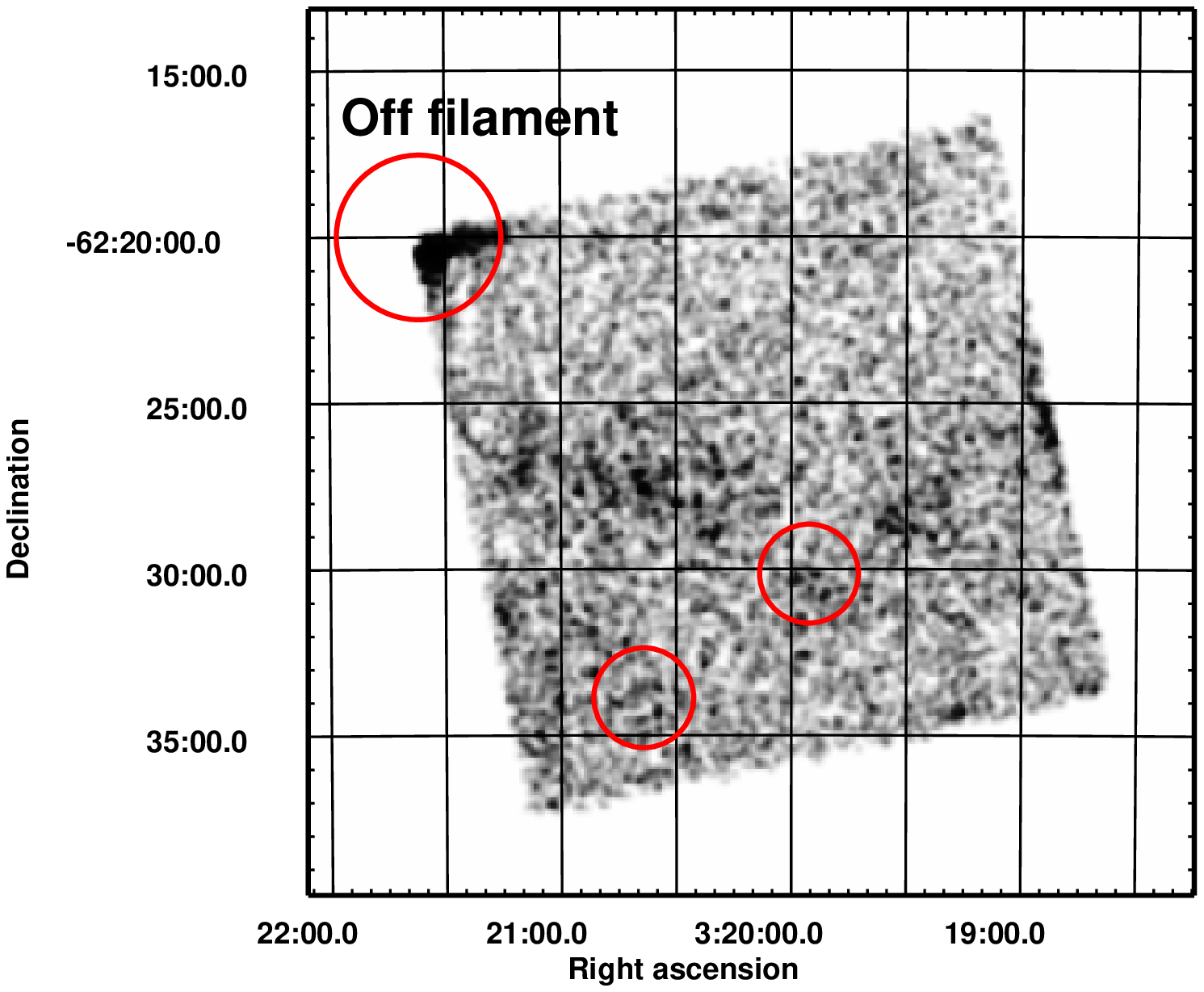}
\caption{Cleaned and smoothed \suzaku\ XIS1 images in the 0.3--5~\kev\ band for our on-filament (\textit{left}) and off-filament (\textit{right}) observations. The data have been binned up by a
factor of 4 in the detector's $x$ and $y$ directions, and then smoothed with a Gaussian whose standard deviation is equal to 1.5 times the binned
pixel size. The particle background has not been subtracted from the data.
The red circles outline the regions that were excluded from the analysis (see text for details).
\label{fig:XIS1images}}
\end{figure*}

Our data were initially processed at NASA Goddard Space Flight Center (GSFC) using processing version 1.2.2.3.
We have carried out further processing and filtering, using HEAsoft\footnote{\scriptsize{http://heasarc.gsfc.nasa.gov/lheasoft}} v6.1.2
and CIAO\footnote{\scriptsize{http://cxc.harvard.edu/ciao}} v3.4. We first combined the data taken in the 3$\times$3 and 5$\times$5 observation modes.
We then selected events with grades 0, 2, 3, 4, and 6, and cleaned the data using the standard data selection criteria given in the
\suzaku\ Data Reduction Guide\footnote{\scriptsize{http://suzaku.gsfc.nasa.gov/docs/suzaku/analysis/abc/abc.html}}. We excluded the times
that \suzaku\ passed through the South Atlantic Anomaly (SAA), and also times up to 436~s after passage through the SAA. We also
excluded times when \suzaku's line of sight was elevated less than 10\degr\ above the Earth's limb and/or was less than 20\degr\ from the bright-Earth
terminator. Finally, we excluded times when the cut-off rigidity (COR) was less than 8~GV. This is a stricter
criterion than that in the Data Reduction Guide, which recommends excluding times with $\mbox{COR} < 6$~GV. However, the higher
COR threshold helps reduce the particle background, and for observations of the SXRB one desires as low a particle background
as possible. The COR threshold that we use has been used for other \suzaku\ observations of the SXRB \citep{fujimoto07,smith07a}.
Finally, we binned the 2.5--8.5~\kev\ data into 256-s time bins, and used the CIAO script \texttt{analyze\_ltcrv.sl} to remove times whose
count-rates differ from the mean by more than 3$\sigma$. The resulting cleaned XIS1 images are shown in Figure~\ref{fig:XIS1images}.

We found that automated point source detection software did not work well on our \suzaku\ images, possibly because of \suzaku's rather
broad point spread function (half-power diameter $\sim 2\arcmin$; \citealp{mitsuda07}).
Therefore, to remove sources which could contaminate our SXRB spectra, we used data from the pre-release of the second \xmm\ Serendipitous Source
Catalogue\footnote{\scriptsize{http://xmmssc-www.star.le.ac.uk/newpages/xcat\_public\_2xmmp.html}} (2XMMp) to determine the locations of sources in our \suzaku\
fields of view. We excluded sources in the catalog with 0.2--4.5~\kev\ fluxes above $5 \times 10^{-14}$ \flux. For sources with fluxes in
the range 5--$30 \times 10^{-14}$ \flux\ we excluded regions of the detector that were within 1.5\arcmin\ of the source. For brighter sources
we excluded regions within 2.5\arcmin\ of the source. The excluded regions are shown in Figure~\ref{fig:XIS1images}.

We extracted spectra from the full XIS1 field of view, excluding the above-mentioned sources. This field of view includes the parts of the chip
illuminated by the Fe-55 calibration source. However, this will not affect our analysis, as we are only interested in X-ray energies
below those of the lines produced by the calibration source. In our analysis we ignore channels 500--504 (1.825--1.84325~\kev),
as there is artificial structure in the spectra at these energies\footnote{\scriptsize{http://suzaku.gsfc.nasa.gov/docs/suzaku/processing/v1223.html}},
due to a non-continuous conversion from pulse height amplitude (PHA) values to pulse invariant (PI) values near the Si~K
edge\footnote{\scriptsize{http://heasarc.gsfc.nasa.gov/docs/suzaku/about/ug2/dotani.pdf}}.
We binned up our spectra so that there were at least 25 counts per bin.

We constructed a particle background spectrum for each observation using tools available from the \suzaku\ Guest Observer
Facility at GSFC\footnote{\scriptsize{http://suzaku.gsfc.nasa.gov/docs/suzaku/analysis/xisbgd0.html}}. These tools generate a
background spectrum from night-Earth observations with the same distribution of cut-off rigidities as our observations.
The background spectra were subtracted from the corresponding source spectra. Within the energy range that we analyzed, there
are instrumental lines in the background spectra at 1.49, 1.74, and 2.12~\kev, due to Al, Si, and Au, respectively.
We found that these lines were not always accurately subtracted from our source spectra, leading to spurious features in the
background-subtracted spectra. We therefore removed these three lines from the background spectra, and interpolated the
surrounding background continuum across this gap. In order to model these lines, we included three extra Gaussians
in the models we fitted to our spectra.

We generated redistribution matrix files (RMFs) using the tool \texttt{mkxisrmf}, and ancillary response files (ARFs) using
the tool \texttt{xissimarfgen} \citep{ishisaki07}. This latter tool takes into account the spatially varying contamination on
the optical blocking filters of the XIS sensors, which reduces the detector efficiency at low energies \citep{koyama07}.
For the ARF calculation we assumed a uniform source of radius 20\arcmin. When generating the spectral response files,
we used the set of calibration database (CALDB) files released on 2007 Jan 31.

\section{\textit{SUZAKU} SPECTRAL ANALYSIS}
\label{sec:SuzakuAnalysis}

\subsection{Description of Model}
\label{subsec:SuzakuModel}

We fit a model to our spectra which consists of components corresponding to the LB, the Galactic halo, and the
extragalactic background due to unresolved active galactic nuclei. For the LB we used a single thermal plasma
model in collisional ionization equilibrium (CIE), while for the Galactic halo we used two equilibrium thermal plasma
components. The use of a two-temperature halo model is well established from analyses of the \rosat\ All-Sky
Survey \citep{kuntz00,snowden00}, and is required to get a good fit to our data (see \S\ref{subsec:SuzakuResults}, below).
For the extragalactic background we used a power-law whose photon index was frozen at 1.46 \citep*{chen97}. The LB
component was not subject to any absorption, whereas the remaining components were. The temperatures of the thermal
emission components and the normalizations of all four emission components were free parameters.

We fit the above-described model simultaneously to our on- and off-filament 0.3--5.5~\kev\ \suzaku\ spectra. Except for the adopted
absorbing column densities, all of the model parameters were assumed to be the same for both spectra. The
difference in absorbing column helps us to separate out the foreground emission (due to the LB) from the background
emission (due to the halo and the extragalactic background). As in \citet{henley07}, we obtained the absorbing column densities
for our observation directions from the \iras\ 100-micron intensities for these directions 
(7.10 [on] and 1.22 [off] MJy sr$^{-1}$; \citealp*{schlegel98}). These were converted to column densities using the conversion relation in \citet{snowden00}.
The resulting on- and off-filament column densities are $9.6 \times 10^{20}$ and $1.9 \times 10^{20}$~\pcmsq, respectively.

We also include data from the \rosat\ All-Sky Survey in our fit \citep{snowden97}. The R1 and R2 count-rates help constrain the model
al lower energies (below $\sim$0.3~\kev), while the higher channels overlap in energy with our \suzaku\ spectra. We extracted the
\rosat\ spectra from 0.5\degr\ radius circles centered on our two \suzaku\ pointing directions using the HEASARC X-ray Background
Tool\footnote{\scriptsize{http://heasarc.gsfc.nasa.gov/cgi-bin/Tools/xraybg/xraybg.pl}} v2.3.

The spectral analysis was carried out using XSPEC\footnote{\scriptsize{http://heasarc.gsfc.nasa.gov/docs/xanadu/xspec/xspec11}}
v11.3.2 \citep{arnaud96}. For the thermal plasma components, we used the Astrophysical Plasma Emission Code (APEC) v1.3.1 \citep{smith01a}
for the \suzaku\ data and the \rosat\ R4--7 bands, and the Raymond-Smith code (\citealt{raymond77} and updates) for the \rosat\ R1--3 bands.
For a given model component (i.e., the LB or one of the two halo components), the temperature and normalization of the \rosat\ Raymond-Smith model
are tied to those of the corresponding \suzaku\ APEC model. We chose to use the Raymond-Smith code for the lower-energy \rosat\ channels
because APEC's spectral calculations below 0.25~\kev\ are inaccurate, due to a lack of data on transitions from L-shell ions of Ne, Mg,
Al, Si, S, Ar, and Ca\footnote{\scriptsize{http://cxc.harvard.edu/atomdb/issues\_caveats.html}}. As the upper-limit of the \rosat\ R1 and R2 bands is 0.284~\kev,
and the R3 band also includes such low-energy photons \citep{snowden97}, APEC is not ideal for fitting to these energy bands.

For the absorption, we used the XSPEC \texttt{phabs} model, which uses cross-sections from \citet{balucinska92}, with
an updated He cross-section from \citet*{yan98}. Following \citet{henley07}, we used the interstellar chemical
abundance table from \citet*{wilms00}. For many astrophysically abundant elements, these abundances are lower than those in
the widely used solar abundance table of \citet{anders89}. However, recently several elements' solar photospheric abundances have been
revised downwards \citep{asplund05a}, and are in good agreement with the \citet{wilms00} abundances. Therefore, like \citet{henley07}
we take the \citet{wilms00} interstellar abundances to be synonymous with solar abundances.

As noted in \S\ref{sec:DataReduction}, we included three Gaussians to model the \suzaku\ instrumental lines from Al,
Si, and Au. The parameters of these lines were independent for the two \suzaku\ observations.

During the course of the spectral modeling, it became apparent that there is a discrepancy in the normalization of the extragalactic power-law (EPL)
between the two \suzaku\ spectra. For a photon index of 1.46, the EPL normalizations for the 2.0--5.5~\kev\ energy range
are $\sim$11 (on-filament) and $\sim$8 (off) \pownorm\ at 1~\kev\ (cf. the expected value is $\sim$10 \pownorm; e.g., \citealp{chen97}). We therefore allowed
the normalization of the EPL to differ for the on- and off-filament spectra; however, the normalizations of the thermal plasma components were
still constrained to be equal for the on- and off-filament spectra. We experimented with other methods for dealing with this discrepancy, but none
of the results were significantly affected. We discuss this discrepancy further in \S\ref{subsec:ExgalNorm}.

\subsection{Results}
\label{subsec:SuzakuResults}

\begin{figure*}
\includegraphics[width=0.48\linewidth]{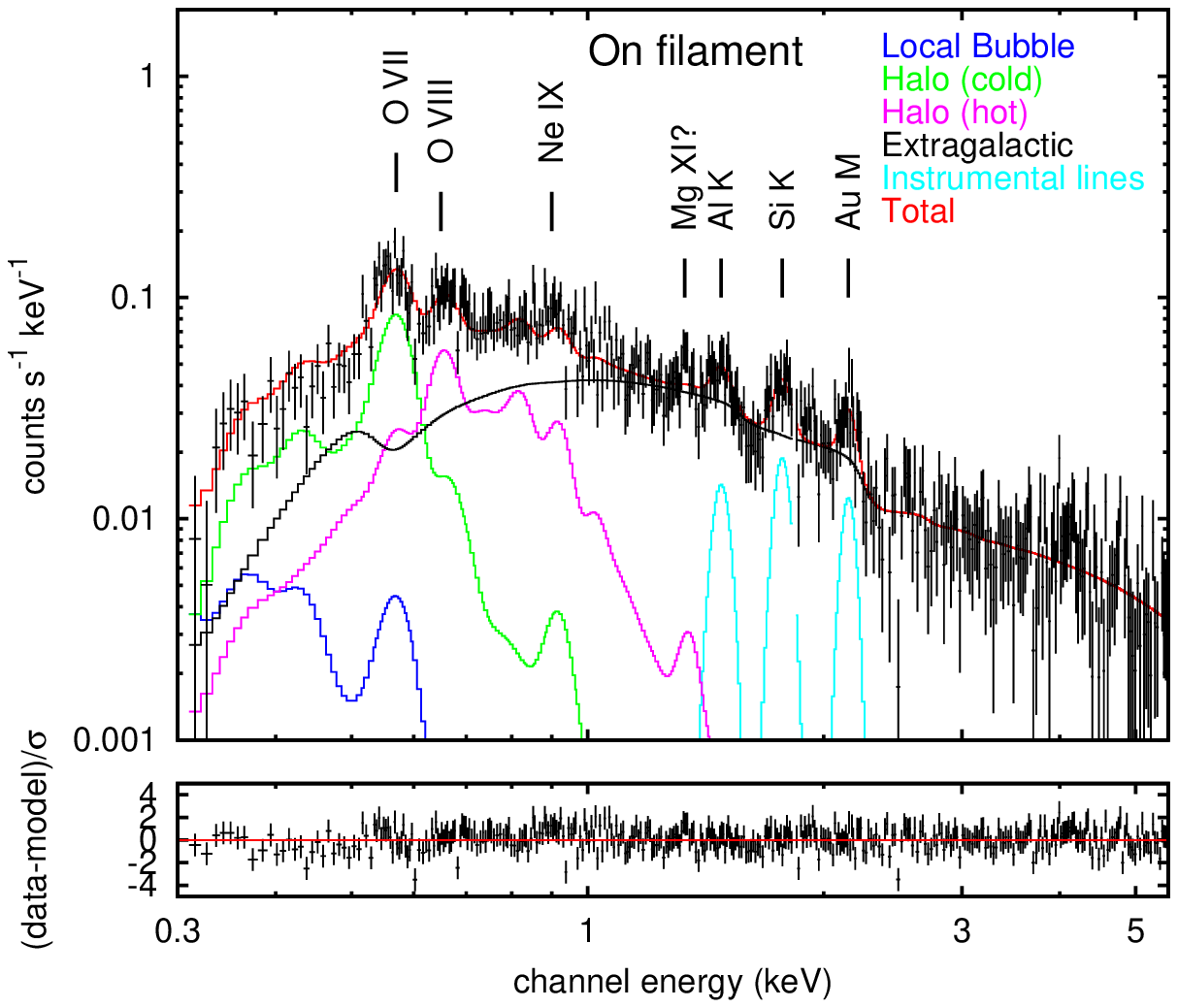}
\hspace{0.04\linewidth}
\includegraphics[width=0.48\linewidth]{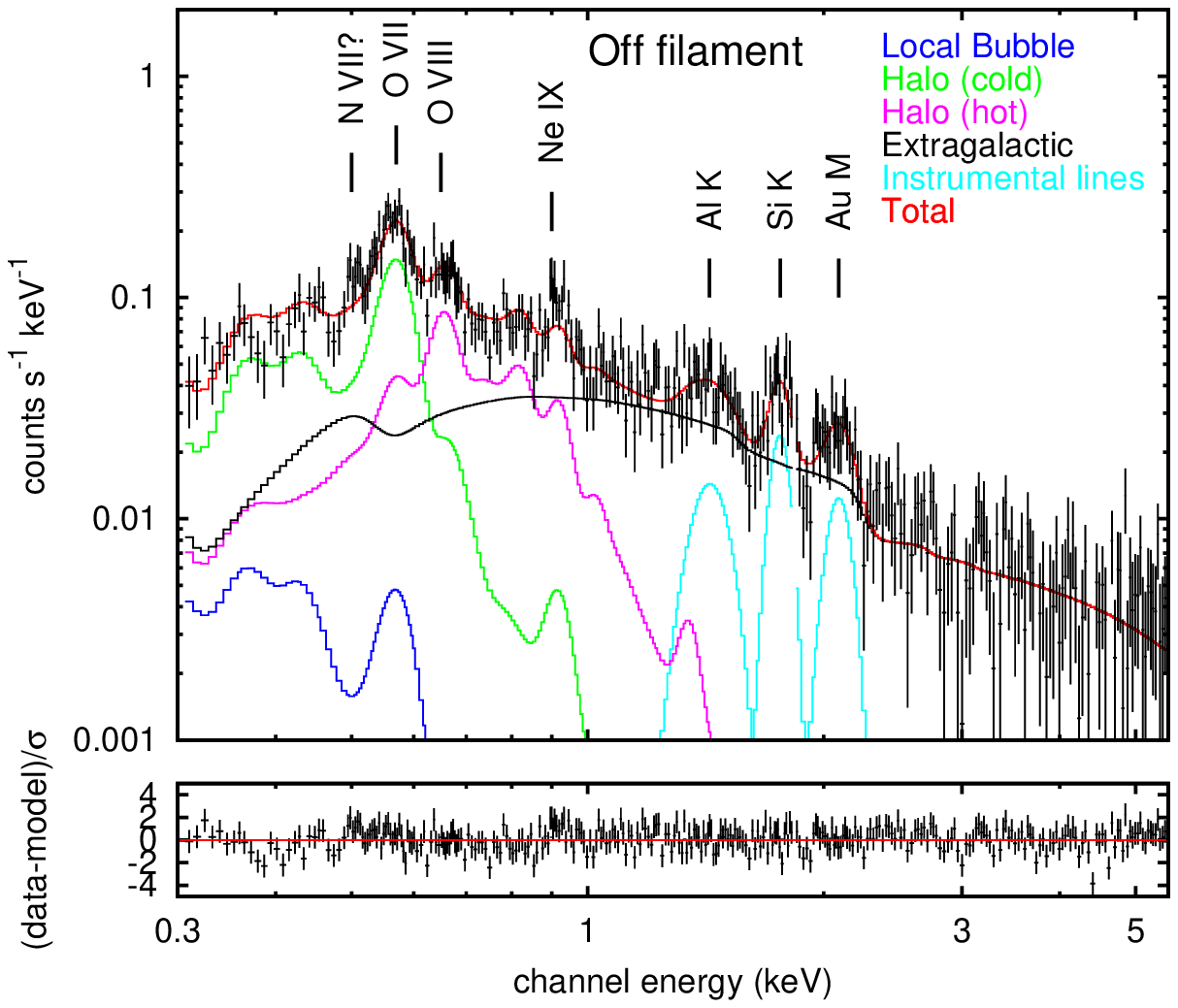}
\caption{Our observed on-filament (\textit{left}) and off-filament (\textit{right}) \suzaku\ spectra, with the best-fitting model obtained by fitting jointly
to the \suzaku\ and \rosat\ data (Model~1 in Table~\ref{tab:FitResults}). The gap in the Si~K instrumental line is where channels 500--504
have been removed from the data (see \S\ref{sec:DataReduction}).
\label{fig:SuzakuSpectra1}}
\end{figure*}

\begin{figure}
\includegraphics[width=0.9\linewidth]{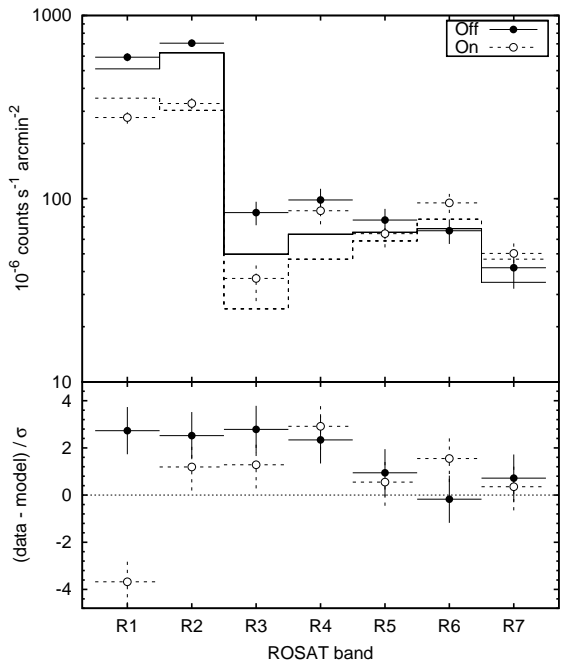}
\caption{The on-filament (\textit{dashed}) and off-filament (\textit{solid}) \rosat\ All-Sky Survey spectra, compared with Model~1 from Table~\ref{tab:FitResults}.
For clarity the individual model components have not been plotted.
\label{fig:ROSATSpectra1}}
\end{figure}

Our on- and off-filament \suzaku\ spectra are shown in Figure~\ref{fig:SuzakuSpectra1}, along with the best-fitting
multicomponent spectral model described in the previous section. The model parameters are presented in Table~\ref{tab:FitResults} (Model~1).
Overall, the model gives a good fit to the data: $\chi^2 = 734.24$ for 703 degrees of freedom. However, the fit to some of the
\rosat\ bands is rather poor, as shown in Figure~\ref{fig:ROSATSpectra1}. This could be due to a discrepancy in the effective area
calibration between \rosat\ and \suzaku. This discrepancy may be due to the uncertainty in the amount of contaminant on the XIS1 optical blocking
filter. For example, if the true amount of contaminant is larger than the amount given by the contamination database (CALDB) contamination model,
the calculated effective area will be larger than the true effective area, and the resulting model normalizations will be too small.

\begin{figure*}
\includegraphics[width=0.48\linewidth]{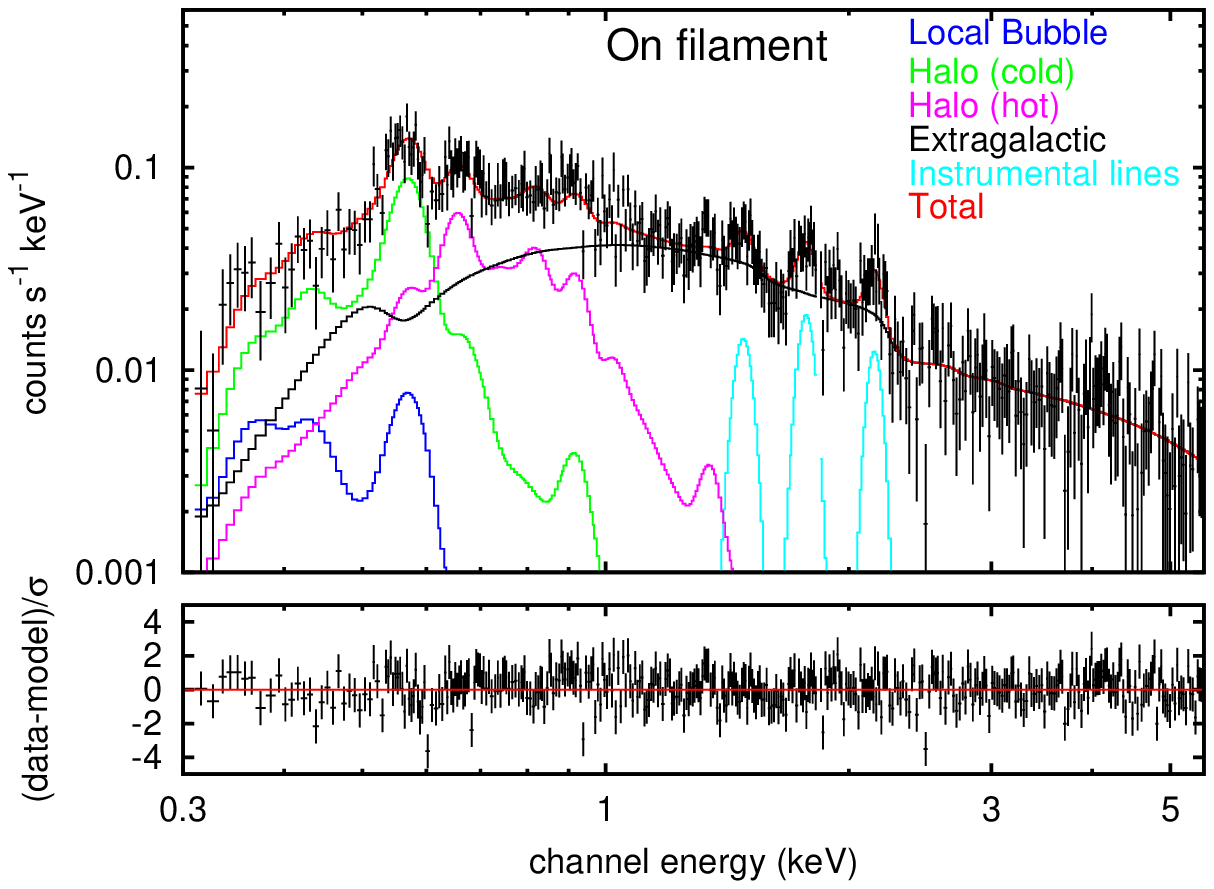}
\hspace{0.04\linewidth}
\includegraphics[width=0.48\linewidth]{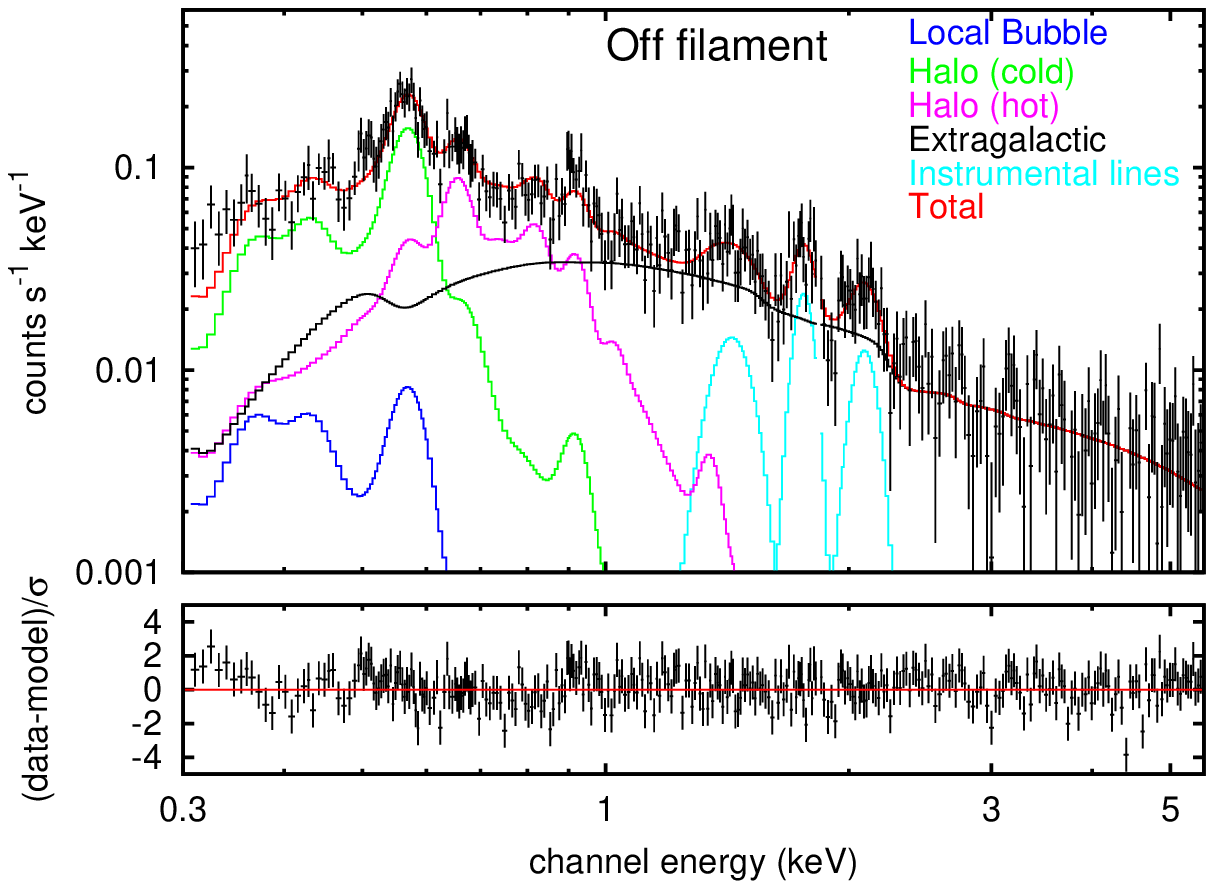}
\caption{As Figure~\ref{fig:SuzakuSpectra1}, but with a \texttt{vphabs} component included for the \suzaku\ spectra to model XIS contamination above that
included in the CALDB (Model~2 in Table~\ref{tab:FitResults}; see \S\ref{subsec:SuzakuResults} for details).
\label{fig:SuzakuSpectra2}}
\end{figure*}

\begin{figure}
\includegraphics[width=0.9\linewidth]{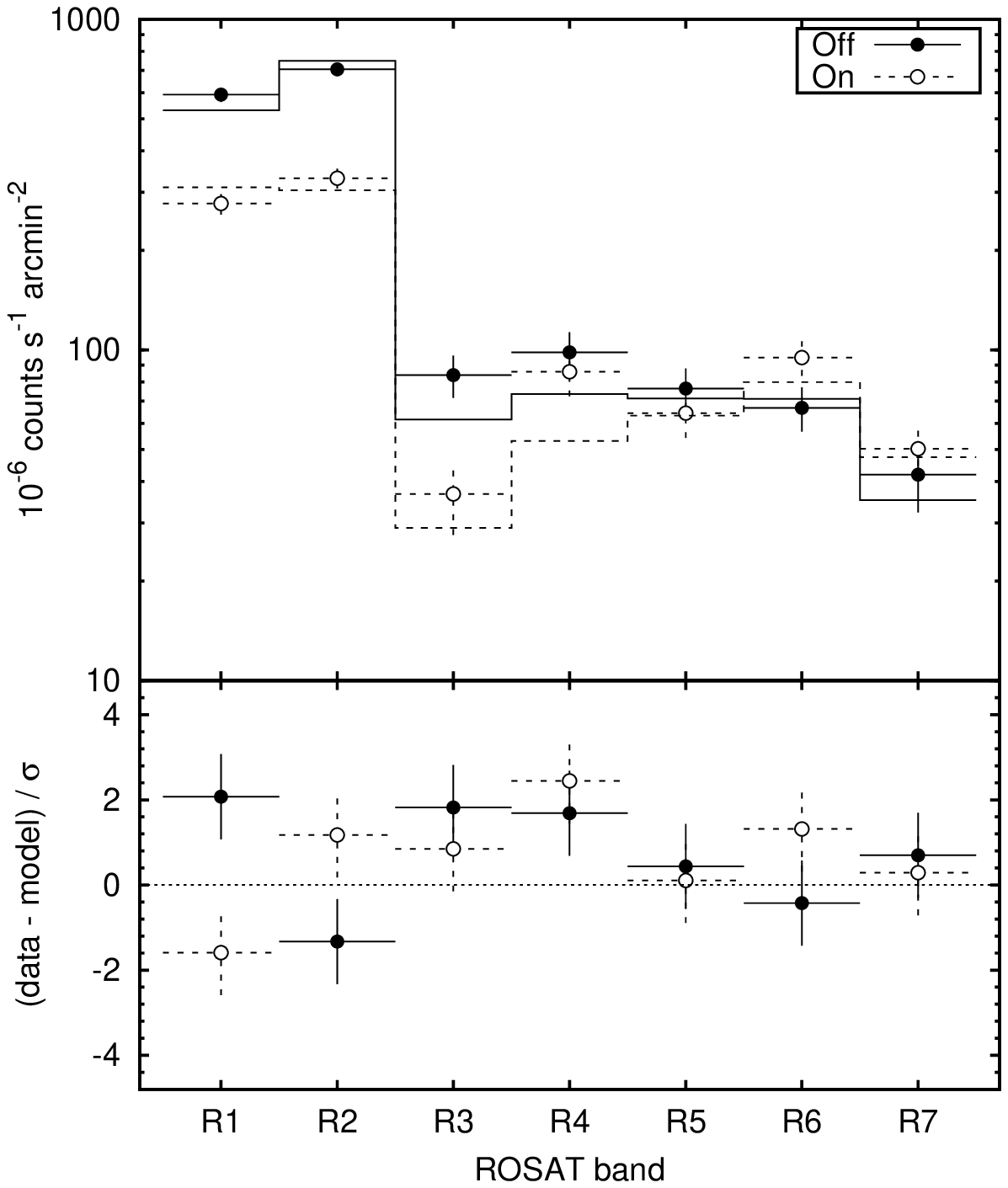}
\caption{As Figure~\ref{fig:SuzakuSpectra2}, but for Model~2 from Table~\ref{tab:FitResults}.
\label{fig:ROSATSpectra2}}
\end{figure}

We investigated this possibility by adding a \texttt{vphabs} absorption component to our model, to model any contamination over and
above that already included in the CALDB. This model component attenuated the emission from the LB, halo, and extragalactic background for
the \suzaku\ spectra only. We adjusted the model oxygen abundance to give $\mathrm{C}/\mathrm{O} = 6$ \citep{koyama07}, and set the
abundances of all other elements to zero. The results of this model are given as Model~2 in Table~\ref{tab:FitResults}.
Figures~\ref{fig:SuzakuSpectra2} and \ref{fig:ROSATSpectra2} show this model compared with the \suzaku\ and \rosat\ spectra, respectively.
One can see from Figure~\ref{fig:ROSATSpectra2} that the fit to the \rosat\ data is greatly improved. The model implies a column
density of carbon atoms, $\NC = (0.28 \pm 0.04) \times 10^{18}$~\pcmsq, in addition to the amount of contamination given by the CALDB contamination
model, which is $\NC = 3.1 \times 10^{18}$~\pcmsq\ at the center of the XIS1 chip (from the CALDB file \texttt{ae\_xi1\_contami\_20061016.fits}).
This correction does not seem unreasonable, given that the systematic uncertainty on the contaminant thickness is
$\sim$$0.5 \times 10^{18}$~\pcmsq.\footnote{\scriptsize{http://www.astro.isas.ac.jp/suzaku/process/caveats/caveats\_xrtxis08.html}}

We investigated whether or not a LB $+$ two-temperature halo model is justified by fitting a model without a LB component to our data, and
by fitting a LB $+$ one-temperature halo model to our data. We also included a power-law to model the extragalactic background.
Both these models gave poor fits to the data: $\chi^2 = 1068.63$ and 870.51, respectively, for 704 degrees of freedom (Models~3 and 4 in Table~\ref{tab:FitResults}).
In both cases, the large values of $\chi^2$ are mainly due to very poor fits to the \rosat\ R1 and R2 data.

Because CIE models give good fits to the data, we did not investigate non-equilibrium ionization models. However, it should be noted that there
are a number of individual features which are not well fit. In the off-filament spectrum, there is excess emission at $\sim$0.5~\kev.
This feature may be too narrow to be an emission line, but if it is it is most likely due to \NVII\ \Lyalpha.
The \NeIX\ \Healpha\ emission at $\sim$0.9~\kev\ is poorly fit in both spectra -- in both spectra
the observed emission is in excess of the model. Finally, there is excess emission in the on-filament spectrum at $\sim$1.3~\kev,
just below the Al~K instrumental line at 1.49~\kev. This is most likely due to \MgXI\ \Healpha. This emission may also be present
in the off-filament spectrum, although it does not show up in the residuals. Note that the Gaussian representing the Al~K line in the
off-filament spectrum is broader and at a lower energy than that in the on-filament spectrum. It is possible that in the off-filament
spectrum the data are not good enough to distinguish clearly the \MgXI\ \Healpha\ line and the instrumental Al~K line, and that
the Gaussian is in fact fitting to both lines, which would both broaden it and lower its energy.

Table~\ref{tab:FitResults} also shows some variants of the above model. In Model~5 we fit exactly the same model to the
\suzaku\ data alone. Without the \rosat\ data, we cannot constrain the LB temperature \TLB. We therefore fix it at the value
determined in Model~2: $\TLB = 10^{5.98}$~K. We also fix $\NC$ for the \texttt{vphabs} contamination component at the Model~2
value: $\NC = 0.28 \times 10^{18}$~\pcmsq. The best-fitting model parameters are in very good agreement with those obtained by
fitting jointly to the \suzaku\ and \rosat\ spectra (compare Models~2 and 5). We can only get consistent results between the \suzaku-\rosat\
joint fit and the fit to just the \suzaku\ data by using a two-temperature model for the halo. However, note that the Model~5 LB emission measure
is consistent (within its errorbar) with zero. This is not to say that our data imply that there is no LB at all: as noted above, we
need a LB component and two halo components to get a good joint fit to the \suzaku\ and \rosat\ data. Instead, the Model~5 results
imply that our \suzaku\ data are consistent with the LB not producing significant emission in the \suzaku\ band (i.e., $E \ga 0.3$~\kev).

Also shown in Table~\ref{tab:FitResults} are the results of fitting our model (without any LB component) to the on- and off-filament
\suzaku\ spectra individually (Models~6 and 7, respectively). As has already been noted, the normalization of the extragalactic background
differs significantly between the two spectra. However, the plasma model parameters for the individual spectra are in good agreement
with each other. These results seem to justify allowing the extragalactic normalization to differ between the two spectra
while keeping all other model parameters equal for the two spectra.

\subsection{Chemical Abundances in the Halo}
\label{subsec:Abundances}

We investigated the chemical abundances of the X-ray-emissive halo gas by repeating the above-described modeling, but allowing the abundances of certain
elements in the halo components to vary. In particular we wished to investigate whether or not varying the neon and magnesium halo abundances
improved the fits to the \NeIX\ and \MgXI\ features noted above. We also allowed the abundance of iron to vary, as iron is an important
contributor of halo line emission to the \suzaku\ band.

For this investigation we just fit to the \suzaku\ data, fixing the LB temperature at $\logTLB = 5.98$, and fixing the carbon column density
of the \texttt{vphabs} contamination component at $\NC = 0.28 \times 10^{18}$~\pcmsq. As we could not accurately determine the level of the continuum
due to hydrogen, we could not measure absolute abundances. Instead, we estimated the abundances relative to oxygen by holding the oxygen abundance
at its \citet{wilms00} value, and allowing the abundances of neon, magnesium, and iron to vary. Both halo components were
constrained to have the same abundances.

The best-fitting temperatures and emission measures of the various model components are presented as Model~8 in Table~\ref{tab:FitResults}, and
the abundances are presented in Table~\ref{tab:Abundances}. The best-fitting model parameters are not significantly affected by 
allowing certain elements' abundances to vary (compare Model~8 with Model~5). Iron
does not seem to be enhanced or depleted relative to oxygen in the halo. Neon and magnesium both appear to be enhanced in the halo relative to
oxygen, which is what one would expect from Figures~\ref{fig:SuzakuSpectra1} and \ref{fig:SuzakuSpectra2}, as the models shown in those figures
underpredict the neon and magnesium emission.

We discuss these results in \S\S\ref{subsec:SuzakuandSWCX} and \S\ref{subsec:Halo}. In \S\ref{subsec:SuzakuandSWCX} we discuss the
possibility that the enhanced neon and magnesium emission is in fact due to SWCX contamination of these lines, rather than being
due to these elements being enhanced in the halo. On the other hand, in \S\ref{subsec:Halo} we discuss the implications of neon
really being enhanced in the halo with respect to oxygen and iron.

\begin{deluxetable}{cc}
\tablecaption{Halo Abundances\label{tab:Abundances}}
\tablehead{
\colhead{Element}	& \colhead{Abundance}
}
\startdata
O			& 1 (fixed)		\\
Ne\tablenotemark{a}	& $1.8 \pm 0.4$		\\
Mg\tablenotemark{a}	& $4.6^{+3.5}_{-2.8}$	\\
Fe			& $1.2^{+0.4}_{-0.5}$	\\
\enddata
\tablecomments{Abundances are relative to the \citet{wilms00} interstellar abundances:
$\mbox{Ne/O} = 0.178$, $\mbox{Mg/O} = 0.051$, $\mbox{Fe/O} = 0.055$.}
\tablenotetext{a}{These enhanced abundances may be an artefact of SWCX contamination; see \S\ref{subsec:SuzakuandSWCX}.}
\end{deluxetable}

\section{COMPARING THE \textit{SUZAKU} AND \textit{XMM-NEWTON} SPECTRA}
\label{sec:XMMAnalysis}

For comparison, Table~\ref{tab:FitResults} also contains the results of the analysis of the \xmm\ spectra from the same observation
directions by \citet{henley07}. The Model~9 results are taken directly from their ``standard'' model. However, it should be noted
that \citet{henley07} used APEC to model all of their data, whereas in the analysis described above we used the
Raymond-Smith code to model the \rosat\ R1--3 bands. We have therefore reanalyzed the \xmm\ + \rosat\ spectra, this time using
the Raymond-Smith code to model the \rosat\ R1--3 bands, and using APEC to model the \rosat\ R4--7 bands and the \xmm\ spectra.
This new analysis allows a fairer comparison of our \xmm\ results with our \suzaku\ results.
The \xmm\ spectra we analyzed are identical to those used by \citet{henley07} -- see that paper for details of the data reduction.
We added a broken power-law to the model to take into account soft-proton contamination in the \xmm\ spectra. This broken power-law was not folded through
\xmm's effective area, and was allowed to differ for the on- and off-filament datasets (see \citealt{henley07}). The presence
of this contamination means we cannot independently constrain the normalization of the extragalactic background. We therefore
freeze the on- and off-filament normalizations at the \suzaku-determined values.

\begin{figure}
\centering
\includegraphics[width=0.9\linewidth, bb=50 50 410 295]{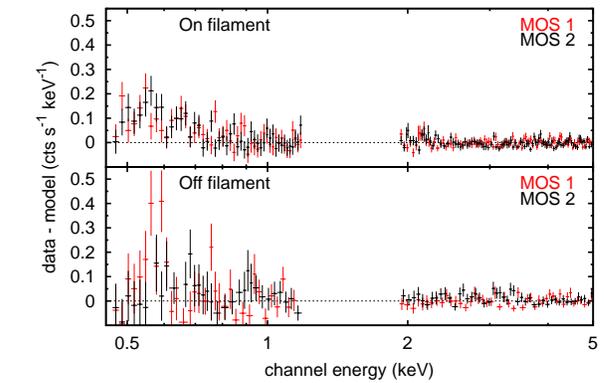}
\caption{The excesses in our on-filament (\textit{top}) and off-filament (\textit{bottom}) \xmm\ spectra over our best-fitting model to the
\suzaku\ + \rosat\ data (Table~\ref{tab:FitResults}, Model~2).
The gap in the data between 1.4 and 1.9~\kev\ is where two bright instrumental lines have been removed.
\label{fig:XMMExcess}}
\end{figure}

The results of this new analysis are presented as Model~10 in Table~\ref{tab:FitResults}. As can be seen, there is poor agreement between
the best-fit parameters of the \suzaku\ + \rosat\ model (Model~2) and the \xmm\ + \rosat\ model (Model~10).
We believe this discrepancy is due to an extra emission component in the \xmm\ spectra.
In Figure~\ref{fig:XMMExcess} we plot the differences between the \xmm\ spectra and our best-fitting \suzaku\ + \rosat\ model
(Model~2 from Table~\ref{tab:FitResults}). To our best-fitting \suzaku\ + \rosat\ model we have added a broken power-law
to model the soft-proton contamination in the \xmm\ spectra. The parameters of this broken power-law are frozen at the values
determined from the fitting to the \xmm\ spectra described in the previous paragraph. The on-filament \xmm\ spectra show
excess line emission at $\sim$0.57 and $\sim$0.65~\kev, most likely due to \OVII\ and \OVIII, respectively. The features in
the off-filament spectra are not as clear. However, there appears to be excess \OVII\ emission in the MOS1 spectrum, and
excess emission at $\sim$0.7~\kev\ (of uncertain origin) and $\sim$0.9~\kev\ (probably \NeIX) in the MOS2 spectrum.
We can estimate the significance of the excess emission by calculating $\chi^2$ for the \xmm\ data compared with the
\suzaku\ + \rosat\ model. We concentrate on the excess oxygen emission and calculate $\chi^2$ for the 0.5--0.7~\kev\
energy range. We find $\chi^2 = 106.28$ for 24 degrees of freedom for the on-filament spectra, and $\chi^2 = 43.03$
for 22 degrees of freedom for the off-filament spectra. These correspond to $\chi^2$ probabilities of
$2.5 \times 10^{-12}$ and 0.0047, respectively, implying that the excesses are significant in both sets of spectra
at the 1\%\ level.

We measure the intensities of the extra oxygen emission by fitting $\delta$-functions at $E = 0.570$~\kev\ and 0.654~\kev\
to the excess spectra in Figure~\ref{fig:XMMExcess}. We fit these $\delta$-functions simultaneously to the on- and off-filament
\xmm\ excess spectra. The intensities of the excess oxygen emission in the
\xmm\ spectra over the best-fitting \suzaku\ + \rosat\ model are $3.8 \pm 0.5$~\LU\ (\OVII) and $1.4 \pm 0.3$~\LU\ (\OVIII).

We believe that this excess oxygen emission is due to SWCX contamination in our \xmm\ spectra. As noted in the Introduction, this was
not taken into account in the original analysis of the \xmm\ spectra. This is because the solar wind flux was steady and slightly
below average during the \xmm\ observations, leading \citet{henley07} to conclude that SWCX contamination was unlikely to be significant.
We discuss the SWCX contamination in our spectra further in \S\ref{sec:SWCX}.

\section{MEASURING THE OXYGEN LINES}
\label{sec:OxygenLines}

As well as using the above-described method to separate the LB emission from the halo emission, we measured the \textit{total}
intensities of the \OVII\ complex and \OVIII\ line at $\sim$0.57 and $\sim$0.65~\kev\ in each spectrum. These lines are a major
component of the SXRB, accounting for the majority of the observed \rosat\ R4 diffuse background that is
not due to resolved extragalactic discrete sources \citep{mccammon02}, and are easily the most prominent lines in our \suzaku\ spectra.

To measure the oxygen line intensities, we used a model consisting of an absorbed power-law, an absorbed APEC model whose oxygen
abundance is set to zero, and two $\delta$-functions to model the oxygen lines. As in the previous section, the power-law models the
extragalactic background, and its photon index was frozen at 1.46 \citep{chen97}. The APEC model, meanwhile, models the line emission
from elements other than oxygen, and the thermal continuum emission. The absorbing columns used were the same as those used in the earlier \suzaku\ analysis.
As with our earlier analysis, we multiplied the whole model by a \texttt{vphabs} component to model the contamination on the optical
blocking filter which is in addition to that included in the CALDB contamination model (see \S\ref{subsec:SuzakuResults}).
We fix the carbon column density of this component at $0.28 \times 10^{18}$~\pcmsq\ (Table~\ref{tab:FitResults}, Model~2).
We fit this model simultaneously to our on- and off-filament spectra. However, all the model parameters were independent for the two directions,
except for the oxygen line energies -- these were free parameters in the fit, but were constrained to be the same in the on- and off-filament spectra.
We used essentially the same method to measure the oxygen line intensities in our \xmm\ spectra. However, we did not use a \texttt{vphabs} contamination
model, and, as before, we added a broken power-law to model the soft-proton contamination.
Table~\ref{tab:OxygenResults} gives the energies and \textit{total} observed intensities of the \OVII\ and \OVIII\ emission measured from our \suzaku\ and 
\xmm\ spectra.

\begin{deluxetable*}{llccccc}
\tablecaption{Observed Total Oxygen Line Intensities\label{tab:OxygenResults}}
\tablehead{
			&			& \multicolumn{2}{c}{\OVII}				&	& \multicolumn{2}{c}{\OVIII}				\\
\cline{3-4} \cline{6-7}
			&			& \colhead{Energy}	& \colhead{Intensity}		&	& \colhead{Energy}	& \colhead{Intensity}		\\
\colhead{Satellite}	& \colhead{Dataset}	& \colhead{(\kev)}	& \colhead{(\LU)}		&	& \colhead{(\kev)}	& \colhead{(\LU)} 		\\
}
\startdata
\suzaku\		& On filament		& 0.564			& $6.51^{+0.37}_{-0.45}$	&	& 0.658			& $2.54^{+0.22}_{-0.28}$	\\
			& Off filament		&			& $10.53^{+0.68}_{-0.55}$	&	&			& $3.21^{+0.25}_{-0.36}$	\\
\xmm\			& On filament		& 0.568			& $10.65^{+0.77}_{-0.82}$	&	& 0.656			& $3.91^{+0.32}_{-0.20}$	\\
			& Off filament		&			& $13.86^{+1.58}_{-1.49}$	&	&			& $2.81^{+0.58}_{-0.60}$	\\
\enddata
\tablecomments{$\LU = \lineunit$.}
\end{deluxetable*}

We can use the difference in the absorbing column for the on- and off-filament directions to decompose the observed line intensities into foreground (LB + SWCX) and
background (halo) intensities. If $I_\mathrm{fg}$ and $I_\mathrm{halo}$ are the intrinsic foreground and halo line intensities, respectively, then the observed on-filament
intensity $I_\mathrm{on}$ is given by
\begin{equation}
	I_\mathrm{on} =I_\mathrm{fg} + \e^{-\tau_\mathrm{on}} I_\mathrm{halo},
\label{eq:I_on}
\end{equation}
where $\tau_\mathrm{on}$ is the on-filament optical depth at the energy of the line. There is a similar expression for the observed off-filament intensity $I_\mathrm{off}$,
involving the off-filament optical depth $\tau_\mathrm{off}$. These expressions can be rearranged to give
\begin{eqnarray}
	I_\mathrm{fg}   &=& \frac{\e^{\tau_\mathrm{on}} I_\mathrm{on} - \e^{\tau_\mathrm{off}} I_\mathrm{off}}{\e^{\tau_\mathrm{on}} - \e^{\tau_\mathrm{off}}}, \\
	I_\mathrm{halo} &=& \frac{I_\mathrm{on} - I_\mathrm{off}}{\e^{-\tau_\mathrm{on}} - \e^{-\tau_\mathrm{off}}}.
\end{eqnarray}
For the purposes of this decomposition, we use the \citet{balucinska92} cross-sections (with an updated He cross-section;  \citealp{yan98}) with
the \citet{wilms00} interstellar abundances. We use the cross-sections at the measured energies of the lines. For the \suzaku\ measurements, the
cross-sections we use are $7.17 \times 10^{-22}$~\cmsq\ for \OVII\ ($E = 0.564~\kev$) and $4.66 \times 10^{-22}$~\cmsq\ for \OVIII\ ($E = 0.658~\kev$).
For the \xmm\ \OVII\ emission we use a cross-section of $7.03 \times 10^{-22}$~\cmsq\ ($E = 0.568$~\kev). We cannot decompose the \xmm\ \OVIII\
emission because the on-filament \OVIII\ line is brighter than the off-filament line. This gives rise to a negative halo intensity, which is unphysical.

The results of this decomposition are presented in Table~\ref{tab:OxygenFGandBG}. Note that the foreground oxygen intensities measured from the \suzaku\
spectra are consistent with zero. This is consistent with our earlier finding that the \suzaku\ spectra are consistent with there being no local emission
in the \suzaku\ band (see \S\ref{subsec:SuzakuResults}). The difference between the foreground \OVII\ intensities measured from our \xmm\ and \suzaku\ spectra
is $5.1 \pm 3.1$~\LU. This is consistent with the \OVII\ intensity measured from the excess \xmm\ emission over the best-fitting \suzaku\ + \rosat\ model
($3.8 \pm 0.5$~\LU; see \S\ref{sec:XMMAnalysis}). The halo \OVII\ intensities measured from our \xmm\ and \suzaku\ spectra are consistent with each other.
This is as expected, as we would not expect the halo intensity to significantly change in $\sim$4 years.

\begin{deluxetable}{lccccc}
\tablecaption{Foreground and Halo Oxygen Line Intensities\label{tab:OxygenFGandBG}}
\tablehead{
			& \multicolumn{2}{c}{\OVII}					&& \multicolumn{2}{c}{\OVIII}			\\
\cline{2-3} \cline{5-6}
			& \colhead{$I_\mathrm{fg}$}	& \colhead{$I_\mathrm{halo}$}	&& \colhead{$I_\mathrm{fg}$}	& \colhead{$I_\mathrm{halo}$}	\\
\colhead{Satellite}	& \colhead{(\LU)}		& \colhead{(\LU)}		&& \colhead{(\LU)}		& \colhead{(\LU)} 		\\
}
\startdata
\suzaku\		& $1.1^{+1.1}_{-1.4}$		& $10.9^{+2.2}_{-1.8}$		&& $1.0 \pm 1.1$		& $2.4^{+1.4}_{-1.5}$		\\
\xmm\			& $6.2^{+2.8}_{-2.9}$		& $8.8^{+4.9}_{-4.6}$		&& \nodata			& \nodata			\\
\enddata
\tablecomments{$\LU = \lineunit$.}
\end{deluxetable}

\section{MODELING THE SOLAR WIND CHARGE EXCHANGE EMISSION}
\label{sec:SWCX}

In \S\S\ref{sec:XMMAnalysis} and \ref{sec:OxygenLines} we presented evidence that our \xmm\ spectra contain an extra
emission component, in addition to the components needed to explain the \suzaku\ spectra. In particular, the \OVII\ and \OVIII\ emission are enhanced in the
\xmm\ spectra. We attribute this extra component to SWCX emission, as it seems unlikely to be due to a change in the Local Bubble or halo
emission. This extra component helps explain why our \xmm\ and \suzaku\ analyses give such different results in Table~\ref{tab:FitResults}.

Previous observations of SWCX have found that increases in the SWCX emission are associated with enhancements in the solar wind, as measured by \ace.
These enhancements consist of an increase in the proton flux, and may also include a shift in the ionization balance to higher ionization stages \citep{snowden04,fujimoto07}.
In \S\ref{subsec:OurSWCXModel}, we present a simple model for heliospheric and geocoronal SWCX emission, and use contemporaneous 
solar wind data from the \ace\ and \wind\ satellites
to determine whether or not the observed enhancement of the oxygen lines in the \xmm\ spectra is due to differences in the solar wind between
our two sets of observations.

In addition to the variability associated with solar wind enhancements, the heliospheric SWCX intensity is also expected to vary during the solar
cycle, due to the different states of the solar wind at solar maximum and solar minimum \citep{koutroumpa06}. As our two sets of observations were
taken $\sim$4 years apart, at different points in the solar cycle, in \S\ref{subsec:Koutroumpa} we examine whether the SWCX intensity variation
during the solar cycle can explain our observations.

\subsection{A Simple Model for Heliospheric and Geocoronal SWCX Emission}
\label{subsec:OurSWCXModel}

\subsubsection{The Basics}
\label{subsubsec:SWCXTheBasics}

A SWCX line from a $\mathrm{X}^{+n}$ ion of element X results from a charge exchange interaction between a $\mathrm{X}^{+(n+1)}$ ion in the solar wind
and a neutral atom. The intensity of that line therefore depends on the density of $\mathrm{X}^{+(n+1)}$ ions in the solar wind, $n_{\mathrm{X}^{+(n+1)}}$,
and on the density of neutral atoms, \nn. The line intensity, $I$, can be written as \citep{cravens00,wargelin04}
\begin{equation}
	I = \frac{1}{4\pi} \int \sigma y \nn \left( \frac{n_{\mathrm{X}^{+(n+1)}}}{\nX} \right) \left( \frac{\nX}{\nH} \right) \nsw \usw dl,
\label{eq:SWCX}
\end{equation}
where $\sigma$ is the cross-section for a charge exchange reaction between a $\mathrm{X}^{+(n+1)}$ ion and a neutral, $y$ is the yield of the
particular line of interest, $n_{\mathrm{X}^{+(n+1)}} / \nX$ is the ion fraction of the $\mathrm{X}^{+(n+1)}$ ion, $\nX / \nH$ is the abundance of
element X in the solar wind, and \nsw\ and \usw\ are the proton density and speed of the solar wind. If, for example, we wish to calculate the intensity
of the \OVII\ emission at $\sim$0.57~\kev, the ``line'' of interest is the blend of $n = 2 \rightarrow 1$ transitions of \Oplussix\ ions, and the
relevant ion fraction to insert in the integrand in equation~(\ref{eq:SWCX}) is that of \Oplusseven\ ions. As was stated in the Introduction,
there are contributions to the SWCX emission from solar wind ions interacting with neutral interstellar H and He atoms distributed throughout the solar system
(heliospheric emission), and with neutral H atoms in the outer reaches of the Earth's atmosphere (geocoronal emission). For the heliospheric emission,
one must calculate the contributions due to interactions with H and He separately, and add them. This is because the cross-sections and yields for
these interactions are different, as are the densities of H and He in the heliosphere.

In the following, we estimate the intensities of the \OVII\ and \OVIII\ emission due to heliospheric and geocoronal SWCX emission for our \xmm\ and \suzaku\ observations.
For this purpose, we adopt simple models for the density of the neutral atoms, and use contemporaneous solar wind data from the \ace\ and \wind\ satellites to insert in
equation~(\ref{eq:SWCX}). To calculate the \OVII\ intensity, we use $\sigma = 3.40 \times 10^{-15}$ and
$1.80 \times 10^{-15}$~\cmsq\ for \Oplusseven\ interacting via charge exchange with H and He, respectively \citep{koutroumpa06}. The yield of \OVII\ \Kalpha\ emission from
interactions between \Oplusseven\ and He is $y = 0.86$, where $y$ is the sum of the contributions of the resonance, intercombination, and forbidden lines \citep*{krasnopolsky04}.
Yield information is not available for \Oplusseven\ interacting with H. However, by definition $y \le 1$, so we can calculate an upper-limit for the expected SWCX emission by
setting $y = 1$.  To calculate the \OVIII\ intensity, we use $\sigma = 5.65 \times 10^{-15}$ and $2.80 \times 10^{-15}$~\cmsq\ for charge exchange interactions of \Opluseight\
with H and He, respectively \citep{koutroumpa06}. The yield of the \OVIII\ \Lyalpha\ line due to charge exchange with He is 0.65 \citep{krasnopolsky04}. Again, yield
information is not available for \Opluseight\ interacting with H, so we calculate an upper limit on the expected intensity by setting $y=1$.

\subsubsection{Solar Wind Data}

\begin{figure*}
\includegraphics[width=0.48\linewidth,bb=50 100 410 554]{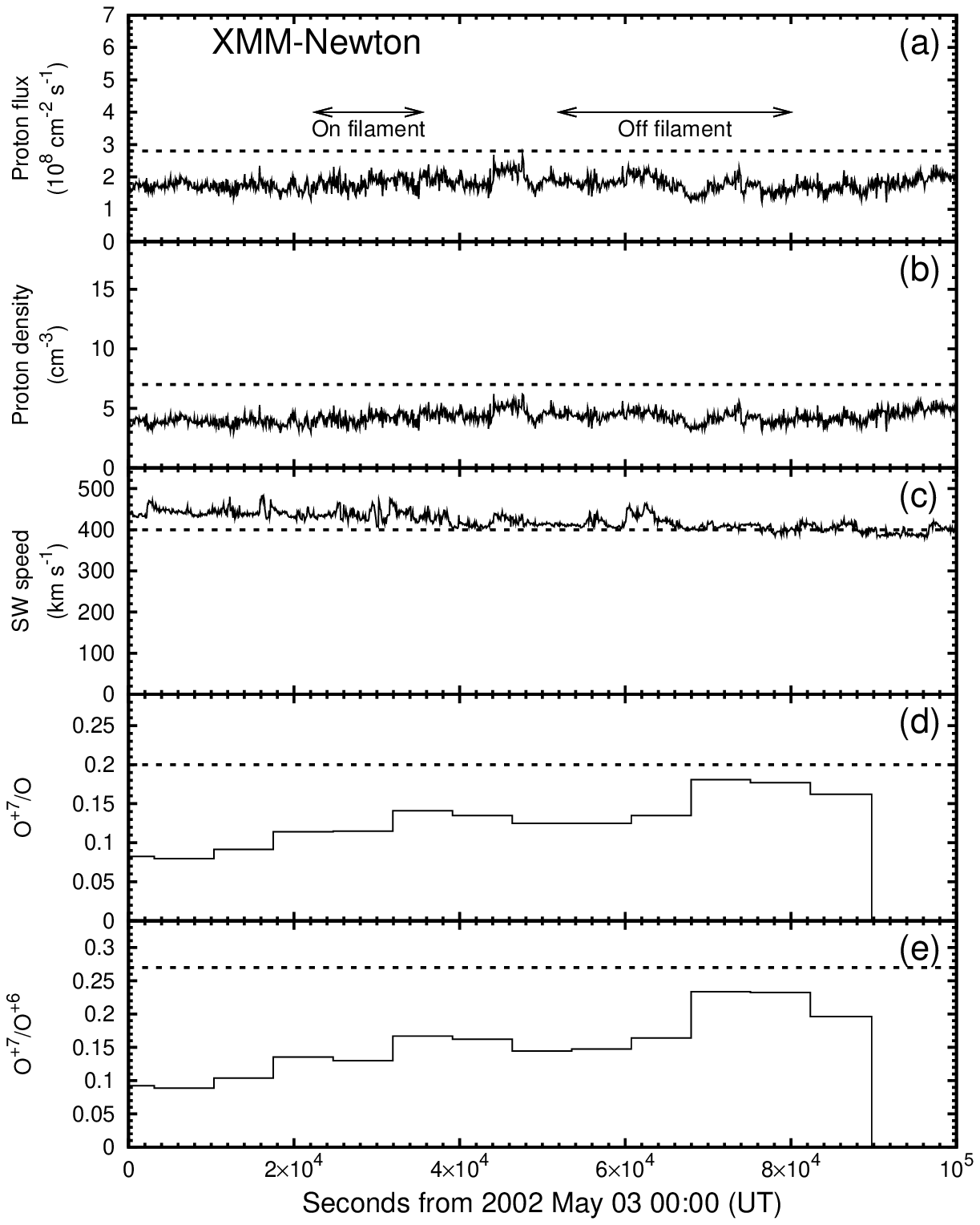}
\hspace{0.04\linewidth}
\includegraphics[width=0.48\linewidth,bb=50 100 410 554]{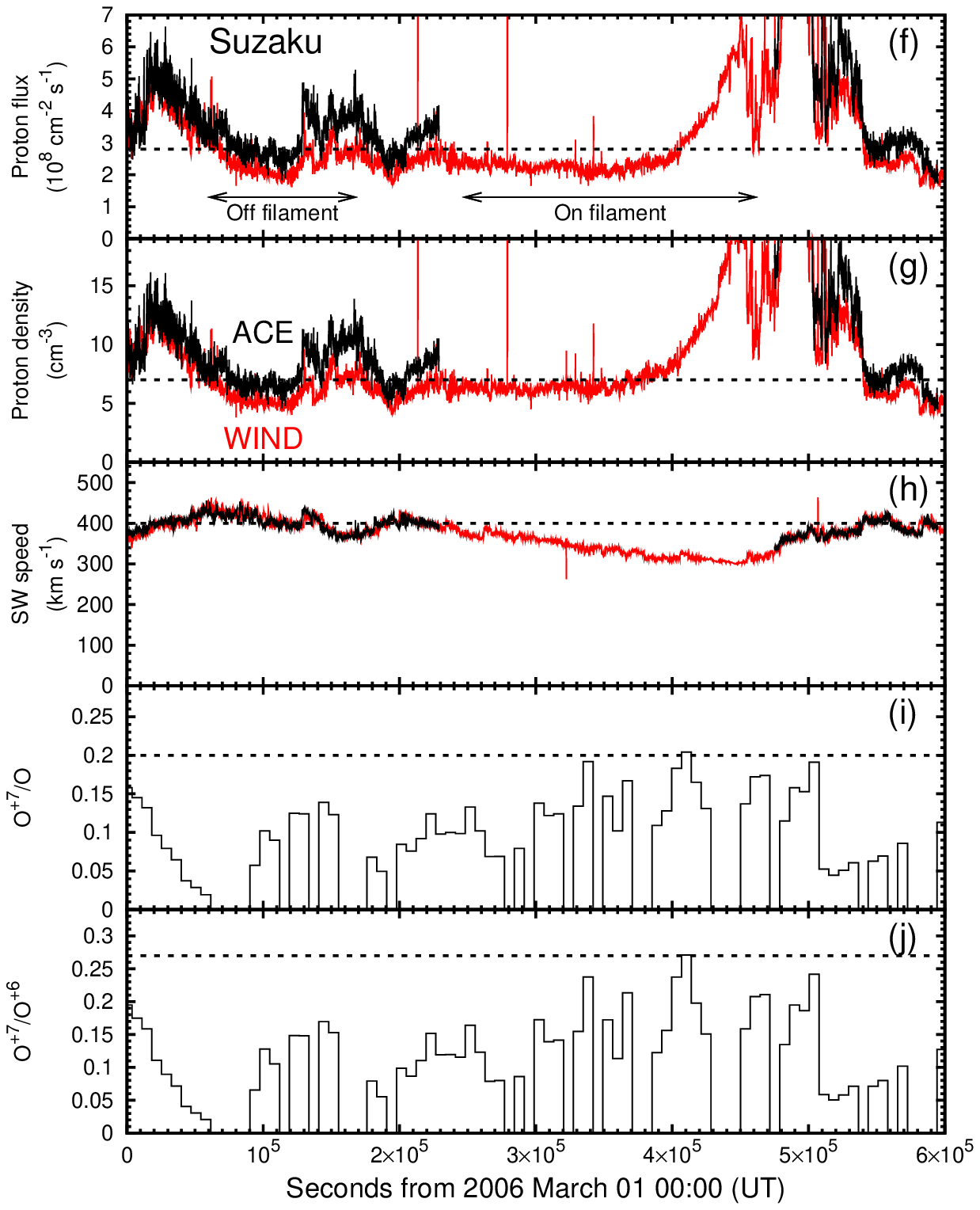}
\caption{Solar wind data from around the times of our \xmm\ (\textit{left}) and \suzaku\ (\textit{right}) observations.
Panels (a) and (f) show the solar wind proton flux, measured by the \ace\ SWEPAM experiment (\textit{black}) and, for the \suzaku\
observations, by the \wind\ SWE experiment as well (\textit{red}). The proton flux is the product of the proton density and the solar wind speed,
shown in panels (b) and (c) for the \xmm\ observations, and panels (g) and (h) for the \suzaku\ observations.
Panels (d) and (i) show the \Oplusseven\ ion fraction, and panels (e) and (j) show the ratio of \Oplusseven\ to \Oplussix. The oxygen ion fraction data
are from the \ace\ SWICS/SWIMS experiment. The black dashed lines show typical values for the slow solar wind, from \citet{wargelin04} and \citet{schwadron00}.
The solar wind data have not been shifted to allow for the travel time from the spacecraft to the Earth ($\sim$4~ks).
\label{fig:SolarWindData}}
\end{figure*}

We use solar wind data from the \ace\ and \wind\ satellites, downloaded from the \ace\ Science Center\footnote{\scriptsize{http://www.srl.caltech.edu/ACE/ASC/level2}} and
CDAWeb\footnote{\scriptsize{http://cdaweb.gsfc.nasa.gov}}, respectively. The solar wind data from around the times of our \xmm\ and \suzaku\ observations are shown
in Figure~\ref{fig:SolarWindData}. Note the different ranges on the time axes -- this is because the \suzaku\ pointings were much longer than the \xmm\
pointings. Where possible, we use the proton density and wind speed (and hence proton flux) from the \ace\ Solar Wind Electron, Proton, and
Alpha Monitor (SWEPAM). In Figure~\ref{fig:SolarWindData}, we just plot good data (bad SWEPAM data are denoted by a value of $-9999.9$ in the density or speed column,
and so cannot be plotted in Fig.~\ref{fig:SolarWindData}). As can be seen in Figures~\ref{fig:SolarWindData}(f--h), there is a gap in the \ace\ SWEPAM dataset
(plotted in black), which is almost exactly coincident with our on-filament \suzaku\ observation. We use data from the \wind\ Solar Wind Experiment (SWE) to
fill in this gap; these data are plotted in red in Figures~\ref{fig:SolarWindData}(f--h). In the \wind\ dataset, good data are denoted by a quality flag of 0.
Unfortunately, for the times covered by Figures~\ref{fig:SolarWindData}(f--h), the \wind\ dataset contains no good data, and hence should not be used.
However, unlike the \ace\ SWEPAM data, densities and speeds are quoted for bad data points, so the data can still be plotted. As can be seen in
Figures~\ref{fig:SolarWindData}(f--h), at times when there is both \wind\ data and \ace\ data, the two satellites are in good agreement regarding the shape of
the solar wind variations. This suggests that the \wind\ data should give a reasonably trustworthy picture of the solar wind proton flux in the gap in the \ace\
data. Figure~\ref{fig:SolarWindData} also shows oxygen ion data from the \ace\ Solar Wind Ion Composition Spectrometer and Solar Wind Ion Mass Spectrometer (SWICS/SWIMS).
The data shown are two-hour averages. Here we only show good data: those with a quality flag of 0.

The \ace\ data from the time of the \xmm\ observations show that the solar wind was very steady at that time, and that the proton flux and \Oplusseven\
ion fraction were slightly below their typical values. As already mentioned, these observations led \citet{henley07} to conclude that SWCX contamination
was likely to be low in the \xmm\ spectra. However, the results of the previous sections imply that this is not the case, and that SWCX emission makes up
most of the foreground oxygen emission in the \xmm\ spectra.

Around the time of the \suzaku\ observations, the solar wind was much more variable. In particular, the proton
flux rises by a factor of $\sim$4 during the last third of the on-filament \suzaku\ observation. However, we find no increase in the soft X-ray (0.3--2.0~\kev)
count-rate associated with this increase in the proton flux. Furthermore, the results of our \suzaku\ spectral fitting do not significantly change if we use
an on-filament spectrum that excludes the final third of the observation, instead of a spectrum from the whole of the on-filament observations.
These results suggest that there are not changes in the SWCX emission directly associated with the changes in the solar wind during the \suzaku\ observations.

From the data shown in Figure~\ref{fig:SolarWindData}, we obtain values of $n_{\mathrm{O}^{+7}} / \nO$, \nsw, and \usw\ to insert into equation~(\ref{eq:SWCX}), in order
to estimate the expected \OVII\ intensity due to SWCX during our \xmm\ and \suzaku\ observations. These values are shown in Table~\ref{tab:SolarWindParameters}. 
The \ace\ data we obtained do not include the \Opluseight\ ion fraction, which is needed to estimate the expected \OVIII\ intensity.
Instead, we use the ratio $\mbox{\Opluseight} / \mbox{\Oplusseven} = 0.35$ from \citet{schwadron00} and our measured \Oplusseven\ ion fractions to estimate the \Opluseight\
ion fraction. In principle we could also obtain $\nO / \nH$ from \ace. In practice this would be measured from the He/O ratio from SWICS/SWIMS. However, these data exhibit
an unexpected 6-month periodicity, meaning that they are unreliable (K.~D. Kuntz, private communication). Instead, we use a canonical slow solar wind oxygen abundance of
$\nO / \nH = 5.6 \times 10^{-4}$ \citep{schwadron00}.

\begin{deluxetable}{lcccc}
\tablecaption{Solar Wind Parameters Used In SWCX Model\label{tab:SolarWindParameters}}
\tablehead{
			& \colhead{${\nsw}_0$\tablenotemark{a}}	& \colhead{$\usw$\tablenotemark{b}}	&								&						\\
\colhead{Satellite}	& \colhead{(\pcc)}			& \colhead{(\kmps)}			& \colhead{$n_{\mathrm{O}^{+7}} / \nO$\tablenotemark{c}}	&  \colhead{$\nO / \nH$\tablenotemark{d}}	\\
}
\startdata
\suzaku\		& 8.3					& 360					& 0.13								& $5.6 \times 10^{-4}$	\\
\xmm\			& 4.2					& 420					& 0.15								& $5.6 \times 10^{-4}$	\\
\enddata
\tablenotetext{a}{Solar wind proton density (from \ace\ and \wind).\\ }
\tablenotetext{b}{Solar wind speed (from \ace\ and \wind).\\ }
\tablenotetext{c}{Solar wind \Oplusseven\ ion fraction (from \ace).\\ }
\tablenotetext{d}{Solar wind oxygen abundance (from \citealp{schwadron00})\\ }
\end{deluxetable}

\subsubsection{Geocoronal Emission}

For the geocoronal emission, we assume that the neutral H density varies with geocentric distance $r_g$ as $\nn(r_g) = {\nn}_0 (10 \RE / r_g)^3$, where
${\nn}_0 = 25$~\pcc\ is the exospheric neutral H density at 10\RE, where \RE\ is the radius of the Earth \citep{wargelin04}. We also assume that the
solar wind ion density is zero inside the magnetosphere, and constant everywhere outside \citep{wargelin04}. With these assumptions, equation~(\ref{eq:SWCX}) yields
\begin{eqnarray}
	I_\mathrm{O\,VII} (\mathrm{geocoronal})	&=& \frac{1}{4\pi} \alpha {\nn}_0 {\nsw}_0 \usw \int^\infty_{\rmin} \left(\frac{10\RE}{r_g}\right)^3 dr_g \nonumber \\
						&=& \frac{\alpha {\nn}_0 {\nsw}_0 \usw}{4\pi} 5\RE \left( \frac{10\RE}{\rmin} \right)^2,
\label{eq:SWCXgeocoronal}
\end{eqnarray}
where $\alpha = \sigma y (n_{\mathrm{O}^{+7}} / \nO) (\nO / \nH)$, ${\nsw}_0$ is the solar wind density at 1~\AU, and \rmin\ is the distance from
the Earth to the magnetopause. For simplicity, we have integrated equation~(\ref{eq:SWCX}) radially away from the Earth, rather than along the line
of sight from the satellite. This approximation is likely to have a bigger effect on the \xmm\ result, as \suzaku\ is in a low-Earth orbit.
In principle, a more accurate intensity could be obtained by integrating equation~(\ref{eq:SWCX}) numerically along the line of sight.
However, as the geocoronal emission is an order of magnitude fainter than the heliospheric emission (see below), the approximation used
in equation~(\ref{eq:SWCXgeocoronal}) is unlikely to affect our conclusions.

For our observations, the angle between the line of sight and the direction toward the Sun was approximately 80\degr. From Figure~7 in \citet{wargelin04},
we estimate that $\rmin \approx 13 \RE$ for this viewing angle. Given the other approximations in the model, a more accurate estimate of \rmin\ is not warranted.

\subsubsection{Heliospheric Emission}
\label{subsubsec:Heliospheric}

To estimate the heliospheric contribution to the SWCX emission, one needs to know the density of neutral H and He throughout the heliosphere.
Neutral H and He are depleted in the inner solar system, due to photoionization and charge exchange \citep{cravens00}.
This effect is stronger for H than for He. Although it is possible to construct detailed models of the density distributions of interstellar neutrals
in the solar system (e.g., \citealp{koutroumpa06}; see \S\ref{subsec:Koutroumpa} below), for this simple estimate we follow \citet{cravens00} and use
$\nn(r_h) = n_{\mathrm{n}0} \exp (-\lambda / r_h)$, where $r_h$ is the heliocentric distance, ${\nn}_0$ is the interstellar density, and
$\lambda$ is the attenuation length. We adopt ${\nn}_0 = 0.15$~\pcc\ and $\lambda = 5$~\AU\ for H, and ${\nn}_0 = 0.015$~\pcc\ and
$\lambda = 1$~\AU\ for He \citep{cravens00}. We also assume that the solar wind is isotropic, so the solar wind density \nsw\ varies as
$\nsw(r_h) = {\nsw}_0 (r_0/r_h)^2$; as before, ${\nsw}_0$ is the solar wind density at $r_h = r_0 = 1~\AU$. We integrate from the
Earth to the edge of the heliosphere. As the distance to the heliopause is much larger than $\lambda$, we can replace the upper limit
with infinity. If, for simplicity, we were to integrate radially away from the Sun, rather than along the line of sight from the Earth,
equation~(\ref{eq:SWCX}) would yield
\begin{eqnarray}
	I_\mathrm{O\,VII} (\mathrm{heliospheric})	&=& \frac{1}{4\pi} \alpha {\nn}_0 {\nsw}_0 \usw \int_{r_0}^\infty \e^{-\lambda / r_h} \left( \frac{r_0}{r_h} \right)^2 dr_h  \nonumber \\ 
							&=& \frac{\alpha {\nn}_0 {\nsw}_0 \usw r_0^2}{4 \pi \lambda} \left( 1 - \e^{-\lambda / r_0} \right),
\label{eq:SWCXheliospheric}
\end{eqnarray}
where $\alpha$ is defined as before. In practice, we integrate equation~(\ref{eq:SWCX}) numerically along the line of sight. This yields
heliospheric intensities that are $\sim$15\%\ larger than those obtained from equation~(\ref{eq:SWCXheliospheric}).

We estimate the intensity of the \OVII\ and \OVIII\ emission due to geocoronal and heliospheric SWCX emission for our \xmm\ and \suzaku\ observations
using the yields and cross-sections from \S\ref{subsubsec:SWCXTheBasics} and the solar wind parameters from Table~\ref{tab:SolarWindParameters}.
The results are shown in Table~\ref{tab:SWCX}.

\begin{deluxetable*}{llccccc}
\tablecaption{Model and Observed SWCX Line Intensities\label{tab:SWCX}}
\tablehead{
			&		& \multicolumn{3}{c}{Model SWCX intensity\tablenotemark{a}}					& \colhead{\citet{koutroumpa07}}			& \colhead{Observed SWCX}\\
\cline{3-5}
			&		& \colhead{Geocoronal}	& \colhead{Heliospheric}	& \colhead{Total\tablenotemark{b}}	& \colhead{model intensity}				& \colhead{intensity}	\\
\colhead{Satellite}	& \colhead{Ion}	& \colhead{(\LU)}	& \colhead{(\LU)}		& \colhead{(\LU)}			& \colhead{(\LU)}					& \colhead{(\LU)}	\\
}
\startdata
\suzaku\		& \OVII\	& 0.28			& 3.4				& 3.7					& 0.83							& $<3.3$\tablenotemark{c}						\\
			& \OVIII\	& 0.16			& 1.9				& 2.0					& 0.07							& $<3.2$\tablenotemark{c}						\\
\xmm\			& \OVII\	& 0.19			& 2.3				& 2.5					& 2.32							& $3.8 \pm 0.5$\tablenotemark{d} -- $7.1 \pm 0.5$\tablenotemark{e}	\\
			& \OVIII\	& 0.11			& 1.3				& 1.4					& 0.92							& $1.4 \pm 0.3$\tablenotemark{d} -- $4.6 \pm 0.3$\tablenotemark{e}	\\
\enddata
\tablecomments{$\LU = \lineunit$.}
\tablenotetext{a}{Calculated using the model discussed in \S\ref{subsec:OurSWCXModel}.\\ }
\tablenotetext{b}{Calculated from the unrounded geocoronal and heliospheric intensities.\\ }
\tablenotetext{c}{$2\sigma$ upper-limit on the foreground oxygen intensity from Table~\ref{tab:OxygenFGandBG}.\\ }
\tablenotetext{d}{Oxygen line intensity measured from the excess of the \xmm\ emission over the best-fitting \suzaku\ + \rosat\ model (see \S\ref{sec:XMMAnalysis}).\\ }
\tablenotetext{e}{Sum of the \xmm-determined lower limit and the \suzaku-determined upper limit.\\ }
\end{deluxetable*}

\subsection{The Heliospheric Emission Model of \citet{koutroumpa06}}
\label{subsec:Koutroumpa}

An alternative approach, developed by \citet{koutroumpa06}, separately considers the contributions of the slow and fast solar winds to the heliospheric SWCX emission,
and how these contributions change with direction and in relation to the solar cycle. They use a more sophisticated, self-consistent model for the distribution of
interstellar neutrals within the solar system, taking into account the effects of gravity, radiation pressure, and losses due to charge exchange and photoionization.
They also calculate self-consistently the distribution of ions in the solar wind, as this too will vary under the effects of charge exchange. Rather than using
solar wind data from \ace\ for a particular time, as we did above, they adopt typical wind velocities, densities, ion fractions, and abundances.

Most importantly for our purposes, \citet{koutroumpa06} calculate heliospheric intensities at solar maximum and at solar minimum. This is relevant to our work because our \xmm\ observations
were taken in 2002 May, near the end of the last solar maximum, whereas our \suzaku\ observations were taken in 2006 March, at solar minimum. At the high ecliptic latitude of
our observation directions ($\approx -74 \degr$), the \citet{koutroumpa06} model predicts higher \OVIII\ intensities at solar maximum than at solar minimum. This is due to a change in the
state of the solar wind. At solar maximum, \citet{koutroumpa06} assume an isotropic ``slow'' solar wind, and at solar minimum they assume a slow solar wind in the solar
equatorial zone, and a ``fast'' solar wind at heliographic latitudes above $+20\degr$ and below $-20\degr$ (the changes in the solar wind during the solar cycle are reviewed
in \citealp{smith03}). The fast solar wind is in a lower ionization state than the slow solar wind \citep{schwadron00}. For example, the ion fractions of \Opluseight\ and
\Oplusseven\ in the fast solar wind are 0\%\ and 3\%, against 7\%\ and 20\%\ for the slow solar wind. These different ion fractions explain why the \citet{koutroumpa06} model
gives \OVIII\ intensities at high ecliptic latitudes that are lower at solar minimum than at solar maximum. At solar minimum a significant length of the line
of sight passes through the fast solar wind, in which there is no \Opluseight\ and so no contribution to the \OVIII\ emission. In contrast, at solar maximum
there are contributions to the heliospheric \OVIII\ emission all along the line of sight to the heliopause.

In a follow-up to \citet{koutroumpa06}, \citet{koutroumpa07} calculate heliospheric oxygen intensity for various \xmm\ and \suzaku\ observations of the SXRB, including
those discussed here. In Table~\ref{tab:SWCX} we give the \OVII\ and \OVIII\ intensities predicted by their ``ground level'' model for our observations.
These ``ground level'' values are calculated assuming the solar wind parameters are at their nominal values, without taking into account possible enhancements in the
solar wind.

\subsection{Comparison with Observations}

As well as the SWCX intensities predicted by the models discussed above, Table~\ref{tab:SWCX} contains the limits on the amount of SWCX oxygen emission observed by
\suzaku\ and \xmm. The upper limits determined from the \suzaku\ spectra are the $2\sigma$ upper limits on the foreground oxygen intensities presented in Table~\ref{tab:OxygenFGandBG}.
We cannot make these upper limits any tighter, as we do not know \textit{a priori} how much of the foreground emission is due to the LB, and how much is due to SWCX.
In \S\ref{sec:XMMAnalysis} we measured the intensities of the oxygen lines in the \xmm\ spectra above the best-fitting \suzaku\ + \rosat\ model.
If we make the reasonable assumption that this excess emission is due to SWCX, the excess oxygen line intensities are lower limits on the SWCX 
oxygen emission observed by \xmm. They are lower limits because we cannot rule out the possibility that our \suzaku\ spectra (which we use as a baseline)
contain some SWCX emission. We can obtain an upper limit on the SWCX oxygen emission observed by \xmm\ by adding to this lower limit the upper limit on the
amount of SWCX emission in the \suzaku\ spectra.

There is poor agreement between the predictions of the model discussed in \S\ref{subsec:OurSWCXModel} and the observed \OVII\ line intensities.
The model overpredicts the \OVII\ intensity observed by \suzaku, and underpredicts that observed by \xmm. As the discrepancies are in opposite senses for the two datasets, they
cannot be reduced by altering input model parameters, such as the cross-sections, line yields, or the attenuation lengths $\lambda$ of neutral H and He
in the heliosphere (see \S\ref{subsubsec:Heliospheric}). This is because the same set of parameters is used to model both datasets, so reducing one
discrepancy will increase the other.

It is likely the model presented in \S\ref{subsec:OurSWCXModel} overpredicts the SWCX \OVII\ intensity for \suzaku\ because we essentially assume an isotropic
slow solar wind when integrating equation~(\ref{eq:SWCX}) to calculate the heliospheric intensity. This is not valid for observations taken at solar minimum
(i.e., the \suzaku\ observations) as much of the line of sight passes through the fast solar wind, which produces less \OVII\ and \OVIII\ emission
because of the lower \Oplusseven\ and \Opluseight\ ion fractions \citep{koutroumpa06}. However, the \citet{koutroumpa06} model does take into account
the state of the solar wind at solar minimum, and one can see from Table~\ref{tab:SWCX} that it predicts lower heliospheric oxygen intensities for
the \suzaku\ observations than our model \citep{koutroumpa07}. The solar minimum oxygen intensities predicted by the \citet{koutroumpa06,koutroumpa07} model
are consistent with the \suzaku\ observations.

As our model assumes an isotropic slow solar wind, the predicted intensities should be more applicable to observations taken at solar maximum
(i.e., the \xmm\ observations). The \OVIII\ intensity predicted by our SWCX model is consistent with the SWCX \OVIII\ intensity
observed by \xmm, while the \citet{koutroumpa06,koutroumpa07} model slightly underpredicts the \OVIII\ intensity. However, both models underpredict the
SWCX \OVII\ intensity observed by \xmm.

The implications of these results are discussed in the following section.

\section{DISCUSSION}
\label{sec:Discussion}

\subsection{Solar Wind Charge Exchange in the XMM-Newton Spectra -- Evidence for a Localized Solar Wind Enhancement}
\label{subsec:SWCXDiscussion}

To summarize the preceding sections, we have found evidence that the oxygen line intensities in our \xmm\ spectra are enhanced
with respect to those in our \suzaku\ spectra. These enhanced intensities are most likely due to SWCX emission. Models for
heliospheric SWCX emission imply that at least part of the difference between the \xmm\ and \suzaku\ spectra can be explain
by a change in the global state of the solar wind between solar maximum and solar minimum \citep{koutroumpa06,koutroumpa07}.
This is because at solar maximum there are more of the high-stage oxygen ions which give rise to heliospheric oxygen emission,
leading to brighter heliospheric oxygen lines.

However, the ``ground level'' \citet{koutroumpa06,koutroumpa07} model, which uses typical values for the various relevant
solar wind parameters, underpredicts the SWCX oxygen intensities measured with \xmm. This implies there is an additional
source of SWCX emission, in addition to that which is generally expected at solar maximum. 
We suggest that the increased oxygen emission during our \xmm\ observations was due to a localized enhancement in the solar wind moving
across the line of sight. This enhancement could have been a region of increased density, or a region of enhanced oxygen abundance.
\citet{koutroumpa07} have also suggested that there was an enhancement in the solar wind during the \xmm\ observations, which
they attribute to a coronal mass ejection starting 2.3~days before the start of the \xmm\ observations.

The ``localized solar wind enhancement'' scenario we have suggested to explain our observations could have important implications
for determining whether or not a SXRB spectrum is likely to be contaminated by SWCX emission.
Previous observations of SWCX emission in the SXRB have found that times of increased SWCX emission appear to be associated with
enhancements in the solar wind as measured by satellites such as \ace\ \citep{snowden04,fujimoto07}. However, in the scenario suggested above,
the solar wind enhancement during the \xmm\ observations was away from the Earth, and so not detectable by \ace. As a result, the
\ace\ data from around the time of the \xmm\ observations show no unusual features (Fig.~\ref{fig:SolarWindData}), which led
\citet{henley07} to conclude that SWCX was unlikely to be significantly contaminating their spectra. Indeed, if contemporaneous
solar wind data are used as an indicator of SWCX contamination, Figure~\ref{fig:SolarWindData}
suggests at first glance that the SWCX emission would be higher during the \suzaku\ observations than during the \xmm\ observations,
which is the exact opposite of what we observe. Our results therefore imply that simply inspecting contemporaneous solar wind data
from \ace\ or other satellites might not be sufficient for determining whether or not a SXRB spectrum is contaminated by SWCX emission.

\subsection{Solar Wind Charge Exchange in the \suzaku\ Spectra?}
\label{subsec:SuzakuandSWCX}

Our analysis suggests that while the \xmm\ spectra are badly contaminated by SWCX emission, particularly in the oxygen lines,
the oxygen lines in the \suzaku\ spectra are not badly contaminated. Other existing \xmm\ and \suzaku\ observations yield oxygen line
intensities due to SWCX of $\sim$5--7~\LU\ for \OVII\ and \OVIII\ \citep{snowden04,fujimoto07}. The upper limits on the SWCX oxygen
emission in our \suzaku\ spectra are much less than these values (see Table~\ref{tab:SWCX}).

However, although the oxygen lines are apparently uncontaminated in our \suzaku\ spectra, there do appear to be other lines in the
spectra which may be due to SWCX (see Fig.~\ref{fig:SuzakuSpectra1}). The off-filament \suzaku\ spectrum shows excess emission at $\sim$0.5~\kev.
For the 6 adjacent channels which make up this feature, $\chi^2 = 14.6$ when compared with our preferred model (Model~2 in Table~\ref{tab:FitResults}).
For 6 degrees of freedom, this corresponds to a $\chi^2$ probability of 0.024, implying the feature is significant at the 5\%\ level.
As noted in \S\ref{subsec:SuzakuResults}, this feature may be too narrow to be an emission line. However, if it is an emission
line, it could be \NVII\ \Lyalpha\ produced by SWCX. The line is unlikely to be due to scattering of solar X-rays off atmospheric
nitrogen, as the line remains in the spectrum even if we exclude times when the elevation of the line of sight above the Earth's
limb is less than 20\degr\ or 25\degr, instead of the default 10\degr\ used in \S\ref{sec:DataReduction}.

The \NeIX\ feature at $\sim$0.9~\kev\ also exhibits excess emission over the best-fitting model -- in this case the excess emission
is seen in both \suzaku\ spectra. This excess emission could be due to an overabundance of neon in the halo (see Table~\ref{tab:Abundances}).
An alternative explanation is that the excess \NeIX\ emission is due to SWCX.

There is also excess emission in the on-filament \suzaku\ spectrum at $\sim$1.3~\kev, which is probably from \MgXI. By fitting a Gaussian
to this feature, we have measured its intensity to be $0.10 \pm 0.07$~\LU, which implies it is only a marginal detection.
As mentioned in \S\ref{subsec:SuzakuResults}, \MgXI\ emission
may also be present in the off-filament \suzaku\ spectrum, where it causes the Gaussian that is modeling the Al~K instrumental line to shift to a
lower energy and increase in width. The \MgXI\ emission could originate from the halo if the magnesium abundance is enhanced by a factor of
$\sim$4 relative to oxygen (see Table~\ref{tab:Abundances}). However, as with neon, an alternative explanation is that the \MgXI\ emission
is due to SWCX. SWCX emission from \MgXI\ has been observed in other \xmm\ and \suzaku\ spectra of the SXRB. \citet{snowden04} found
\MgXI\ emission due to SWCX among a set of \xmm\ spectra of the Hubble Deep Field North, with an intensity of $0.40 \pm 0.08$~\LU, while
\citet{fujimoto07} measured a \MgXI\ SWCX intensity of $0.73^{+0.19}_{-0.20}$~\LU\ from a \suzaku\ spectrum of the North Ecliptic Pole.
These other results imply that the \MgXI\ feature in our on-filament \suzaku\ spectrum is not unusually bright for SWCX emission.

The observation of SWCX emission from \MgXI\ requires the presence of H-like Mg$^{+11}$ ions in the solar wind. Mg$^{+11}$ is the dominant
ionization stage of magnesium in the temperature range $\sim$7--$9 \times 10^6$~K \citep{mazzotta98}. As the solar wind ions
that give rise to SWCX originate in the solar corona, and are frozen into the plasma as it expands away from the Sun \citep{cravens02},
the observation of SWCX emission from \MgXI\ may imply very high temperatures in the solar corona. However, more detailed modeling
is needed to determine if the observed \MgXI\ SWCX intensities imply an unusually high Mg$^{+11}$ ion fraction in the solar wind.
Such modeling of the \MgXI\ and other SWCX lines is beyond the scope of this paper. Nevertheless, the presence of these lines in our
\suzaku\ spectra raises an interesting question: if there is SWCX emission from \MgXI, and possibly \NeIX\ and \NVII, why is there not
significant SWCX emission from \OVII\ and \OVIII?

\subsection{The Normalization of the Extragalactic X-ray Background}
\label{subsec:ExgalNorm}

As was noted in \S\ref{subsec:SuzakuModel}, and as can be seen in Table~\ref{tab:FitResults}, the normalization of the extragalactic background is different
in our two \suzaku\ spectra, with the off-filament value ($7.7 \pm 0.3$ \pownorm\ at 1~\kev) lower than the expected value of $\sim$10 \pownorm\ (e.g., \citealp{chen97}).
One possible reason for this is incorrect subtraction of the particle background: if the particle background spectrum used for our off-filament observation
is too bright, this would result in our background-subtracted SXRB spectrum being fainter than expected. However, we do not think this is likely to be the cause.
At high energies  we do not expect any cosmic X-ray emission, and so any counts in our \suzaku\ spectra at these energies will be
due to the particle background. For each observation, the 10--12 \kev\ count-rates in the non-background-subtracted source spectrum and in the particle background spectrum
(generated from night-Earth data) agree within 8\%. In contrast to this, the particle background flux would have to be reduced by $\sim$30\%\ in order to increase
the off-filament extragalatic normalization to 10.5 \pownorm. It therefore does not seem likely that an incorrect particle background normalization is the reason for this
discrepancy in the on- and off-filament extragalactic normalizations.

As well as allowing the extragalactic normalization to differ between our two \suzaku\ spectra, while keeping everything else the same, we also tried two
other methods for dealing with this discrepancy. First, we simply ignored the discrepancy, and forced the model to be identical for both the on- and
off-filament datasets. Second, we tried multiplying our entire model (Local Bubble + halo + EPL) by a variable factor $f_\mathrm{off}$ when fitting
it to our off-filament \suzaku\ spectrum; this multiplicative factor was an additional free parameter in the fit. The second method essentially assumes
that the discrepancy is due to a reduction of $\sim$20\%\ in \suzaku's sensitivity between the on- and off-filament observations. We do not think this is
plausible, but we still tried out this model.

We found that these two new methods gave temperatures in good agreement with each other and with those in Table~\ref{tab:FitResults}, although the
emission measures of the Local Bubble and of the cool halo component varied by up to $\sim$15\%\ and $\sim$25\%\ between the models, respectively.
Of the three methods we tried, our original method (i.e., that presented in Table~\ref{tab:FitResults}) gave the best fits to the data. Also, when
we fitted our multicomponent model to our on- and off-filament spectra individually, we found that the plasma model components are in good agreement
with each other (Table~\ref{tab:FitResults}, Models~6 and 7). As was noted in \S\ref{subsec:SuzakuResults}, this seems to support our method
for handling the discrepancy in the extragalactic normalization.

In summary, we do not know why the extragalactic normalization differs between our two spectra. It seems unlikely to be due to incorrect
particle-background subtraction, or to a $\sim$20\%\ reduction in \suzaku's sensitivity in the space of a few days. We have tried
several different methods for handling this discrepancy, and the results they give are generally in good agreement with each other. We
therefore conclude that, despite the discrepancy with the extragalactic normalization, the plasma model parameters we have obtained should
be reasonably robust.

\subsection{The Local Bubble}
\label{subsec:LocalBubble}

Because of the SWCX contamination of the oxygen lines in the \xmm\ spectra, and the apparent lack of such contamination in the \suzaku\ spectra,
the best-fitting model of the LB we have derived from our \suzaku\ spectra is quite different from that in \citet{henley07}. In that project,
all the foreground oxygen emission in the \xmm\ spectra was thought to be from the LB, leading to a higher LB temperature than we have obtained
from our \suzaku\ + \rosat\ analysis: $\logTLB = 6.06$, instead of $\logTLB = 5.98$. The lower foreground oxygen intensities in the \suzaku\
spectra cannot be explained by simply lowering the LB emission measure, as the \rosat\ R12 data place a constraint on the LB emission measure.

Note that our LB emission measure ($0.0064 \pm 0.0003$; Table~\ref{tab:FitResults})
is roughly one-third of the value determined by \citet{henley07}. However, this is merely because we used different plasma emission codes
to model the \rosat\ R12 data. We used the Raymond-Smith code to model the R12 data, whereas \citet{henley07} used APEC. For temperatures around
$10^6$~K, the Raymond-Smith code predicts $\sim$3 times as much 1/4-\kev\ flux as APEC. This means that for a given observed flux the
Raymond-Smith-derived emission measure will be $\sim$3 times smaller than the APEC-derived value, which is exactly what we find.
Our LB emission measure is in reasonable agreement with the range of LB emission measures determined from the \rosat\ All-Sky Survey 1/4-\kev\ data
(0.0018--0.0058~\emismeas; \citealp{snowden98}), which was also modeled using the Raymond-Smith code.

The LB temperature that we measure ($\logTLBsq = 5.98^{+0.03}_{-0.04}$; Table~\ref{tab:FitResults}) is lower than the temperatures obtained from
other recent shadowing observations made with \xmm\ and \suzaku: $\logTLB \approx 6.04$--6.06 (\citealp{galeazzi07}, using MBM~20 as a shadow) and
$\approx$6.08 (\citealp{smith07a}, using MBM~12). The range of values from \citet{galeazzi07} are from their results for different
abundance tables \citep{anders82,anders89,grevesse98,lodders03}. Their derived LB temperature is fairly insensitive to the set of
abundances that they used. If we repeat our spectral analysis using \citet{anders89} abundances, we obtain $\logTLB = 6.07 \pm 0.05$, which is
consistent with the \citet{galeazzi07} and \citet{smith07a} temperatures. However, when we use the other abundances tables used by
\citet{galeazzi07}, we obtain LB temperatures of $\logTLB = 5.98^{+0.05}_{-0.04}$ \citep{anders82}, $5.97 \pm 0.04$ \citep{grevesse98},
and $5.96 \pm 0.04$ \citep{lodders03}. These values are closer to our original LB tempeature, obtained using \citet{wilms00} abundances
(which were not used by \citealp{galeazzi07}). Hence, while the LB temperature we obtain using \citet{anders89} abundances is in good
agreement with that of \citet{galeazzi07}, the temperatures obtained using other abundance tables \citep{anders82,grevesse98,lodders03}
are systematically lower than those of \citet{galeazzi07}. These results imply that the LB may not be isothermal.

If we were to assume that the LB is isothermal, it would make it more difficult to explain the lower foreground \OVII\ intensity measured from our \suzaku\ spectra:
$1.1^{+1.1}_{-1.4}$~\LU\ (see Table~\ref{tab:OxygenFGandBG}), compared with $2.63 \pm 0.78$~\LU\ \citep{galeazzi07} and $3.53 \pm 0.26$~\LU\ \citep{smith07a}.
In an isothermal LB, a larger X-ray intensity in a given direction can be explained by the LB being of greater extent in that direction.
Analysis of the \rosat\ All-Sky Survey suggests that the LB radius toward the filament is similar to that toward MBM~20, and the LB radius toward
MBM~12 is roughly two-thirds of this value (\citealp{snowden98}, their Fig.~10). This means that the different observed \OVII\ intensities cannot
be explained by the LB being of different extent in different directions. However, if the LB temperature is higher toward MBM~20 and MBM~12, this
would give rise to brighter \OVII\ emission in these directions.

If we assume the radius of the LB is 100~pc in the direction we have analyzed, our emission measure gives an electron density of 0.0080~\pcc.
Combining this with our LB temperature gives a LB thermal pressure of $\pLB / k = 1.92 \Ne \TLB = \mbox{14,700}$~\presalt\ (13,100--16,100~\presalt).
\citet{galeazzi07} obtained $\pLB / k = \mbox{16,200}$--19,500~\presalt, while \citet{smith07a} obtained $\pLB / k = \mbox{22,000}$~\presalt\
(assuming $\logTLBsq = 6.08$). These results suggest that the LB might not be in pressure equilibrium. Alternatively, they may indicate that
the spectra analyzed by \citet{galeazzi07} and \citet{smith07a} are contaminated by SWCX emission.

\subsection{The Halo}
\label{subsec:Halo}

As we found in \S\ref{subsec:Abundances}, a possible explanation for the excess \NeIX\ emission visible at $\sim$0.9~\kev\ in
Figures~\ref{fig:SuzakuSpectra1} and \ref{fig:SuzakuSpectra2} is that neon is enhanced in the halo, relative to oxygen and iron.
(The alternative is that this emisson is due to SWCX, as mentioned in \S\ref{subsec:SuzakuandSWCX}.) If neon is really enhanced
relative to oxygen and iron in the halo (or, equivalently, if oxygen and iron are depleted relative to neon), this may be evidence
that some of the oxygen and iron in the halo is in dust, rather than in the hot gas. Neon, being chemically inert, does not condense
into dust.

Our best-fitting halo model predicts $\sim$300~\LU\ for the intrinsic \OVI\ $\lambda\lambda$1032,~1038 doublet intensity (using data from ATOMDB).
Note that this value is much smaller than that in \citet{henley07}. This is because their cooler halo component was cooler than ours, and had a
larger emission measure (compare Models~2 and 9 in Table~\ref{tab:FitResults}). Our model \OVI\ intensity is $\sim$20 times smaller than the intrinsic
halo \OVI\ intensity determined by \citet*{shelton07} from an off-filament \textit{Far Ultraviolet Spectroscopic Explorer} (\fuse)
spectrum. This implies that there is both hot, X-ray-emitting gas and warm, ultraviolet-emitting gas in the Galactic halo. \citet{shelton07}
present a model for the distribution of warm and hot gas in the halo, using \OVI\ and \rosat\ data as constraints. Further modeling of the hot
gas distribution in the halo, using the \suzaku\ data discussed here, will be presented elsewhere (S.~J. Lei et~al., in preparation).

\section{SUMMARY \& CONCLUSIONS}
\label{sec:Summary}

We have analyzed a pair of \suzaku\ spectra of the SXRB, obtained from pointings toward and to the side of a nearby
absorbing filament in the southern Galactic hemisphere. We used a shadowing technique to decompose the spectra into their various thermal
emission components, due to the LB and the Galactic halo, using \rosat\ 1/4-\kev\ data to help constrain the model at lower energies.
We find that our best-fitting model parameters (temperatures and emission measures) do not agree with those determined by
\citet{henley07} from a pair of \xmm\ observations in the same directions as our \suzaku\ observations. We attribute this discrepancy
to SWCX contamination in the \xmm\ spectra, particularly in the oxygen lines, which was not taken into account in the original analysis.

Our results suggest that inspecting contemporaneous solar wind data might not be sufficient for determining if a SXRB spectrum is
contaminated by SWCX emission. We have reached this conclusion by comparing our observations with models for SWCX emission.
The simple model of heliospheric and geocoronal SWCX emission that we have presented, which assumes an
isotropic solar wind and uses contemporaneous solar wind data from \ace\ and \wind, cannot explain the different amounts of SWCX
oxygen emission observed by \xmm\ and \suzaku. The more sophisticated heliospheric emission model of \citet{koutroumpa06,koutroumpa07},
which takes into account the change in the global state of the solar wind between solar maximum, can partly explain why more
SWCX emission is observed by \xmm\ than by \suzaku. However, this model underpredicts the amount of oxygen
emission observed by \xmm\ at solar maximum. Instead, the excess SWCX emission in the \xmm\ 
spectra may have been due to a localized enhancement in the solar wind moving across the line of sight during the \xmm\ observations.
If this enhancement was not near the Earth, it would not have shown up in the solar wind data from \ace\ and \wind.

While it appears that the oxygen lines in our \suzaku\ spectra are not badly contaminated by SWCX emission, we cannot
be certain of the true level of SWCX emission in our spectra. Indeed, it is possible that there is SWCX emission from
nitrogen, neon, and magnesium in our \suzaku\ spectra.
However, if in general the level of SWCX contamination is low in our \suzaku\ spectra, as we think is the case, then our
best-fitting models of the LB and halo should be more accurate than those obtained by \citet{henley07}. Our best-fitting LB
temperature is lower than the temperatures obtained from other recent \xmm\ and \suzaku\ observations \citep{galeazzi07,smith07a},
suggesting that the LB is not isothermal. Our LB model also yields a lower pressure than these other LB models, implying that the LB
may not be in pressure equilibrium. Our modeling of the halo emission, meanwhile, suggests that oxygen and iron may be depleted
relative to neon, possibly because they are partly in a dust phase.

\acknowledgements
We would like to thank T.~E. Cravens for an informative discussion on SWCX, B.~J. Wargelin for helpful comments (particularly for suggesting
the idea of a localized enhancement in the solar wind during the \xmm\ observations), and P.~C. Stancil for help in obtaining charge exchange
cross-sections and yields from the literature. We are also grateful to D. Koutroumpa for supplying us with a preprint of \citet{koutroumpa07}.
We would also like to thank K.~D. Kuntz for reducing the \xmm\ data originally analyzed by \citet{henley07},
which we have used again here, and for telling us of his analysis of data from \ace.
We thank the \ace\ SWEPAM and SWICS/SWIMS instrument teams and the \ace\ Science Center for providing the \ace\ data,
and the \wind\ SWE team and the Space Physics Data Facility at Goddard Space Flight Center for providing the \wind\ data.
Finally, we thank the anonymous referee for helpful comments which have greatly improved this paper.
This work was funded by NASA grants NNG04GD78G, awarded through the Long Term Space Astrophysics Program, and NNX07AB03G,
awarded through the \suzaku\ Guest Observer Program.

\bibliography{references}

\clearpage

\begin{landscape}
\begin{deluxetable}{l@{\hspace{-2.5cm}}cccccccccccccc}
\tablecaption{Spectral Fit Results\label{tab:FitResults}}
\tablehead{
		&					& \multicolumn{2}{c}{Local Bubble}				&& \multicolumn{2}{c}{Halo (cool)}				&& \multicolumn{2}{c}{Halo (hot)} 				&& \multicolumn{2}{c}{EPL norm\tablenotemark{a}}	&									\\
\cline{3-4} \cline{6-7} \cline{9-10} \cline{12-13}
		&					& \colhead{$\log T$}	& \colhead{E.M.\tablenotemark{b}}	&& \colhead{$\log T$}	& \colhead{E.M.\tablenotemark{b}}	&& \colhead{$\log T$}	& \colhead{E.M.\tablenotemark{b}}	&& \colhead{On}		& \colhead{Off}			& \colhead{\NC\tablenotemark{d}}					\\
\colhead{Model} & \colhead{Dataset(s)\tablenotemark{c}}	& \colhead{(K)}		& \colhead{($10^{-3}$ \emismeas)}	&& \colhead{(K)}	& \colhead{($10^{-3}$ \emismeas)}	&& \colhead{(K)}	& \colhead{($10^{-3}$ \emismeas)}	&& \colhead{Filament}	& \colhead{Filament}		& \colhead{($10^{18}~\pcmsq$)}		& \colhead{$\chisq/\mbox{dof}$} \\
}
\startdata
1		& S + R					& $5.92^{+0.02}_{-0.04}$& $7.2^{+0.4}_{-0.3}$			&& $6.12 \pm 0.01$	& $24.2^{+1.8}_{-2.3}$			&& $6.50 \pm 0.02$	& $5.7^{+0.4}_{-0.5}$			&& $11.1 \pm 0.3$	& $7.6 \pm 0.3$			& 0\tablenotemark{e}			& $734.24/703$			\\
2		& S + R					& $5.98^{+0.03}_{-0.04}$& $6.4^{+0.3}_{-0.3}$			&& $6.11 \pm 0.01$	& $34.6^{+2.3}_{-2.9}$			&& $6.50^{+0.01}_{-0.02}$& $6.5^{+0.7}_{-0.5}$			&& $11.2 \pm 0.3$	& $7.7 \pm 0.3$			& $0.28 \pm 0.04$			& $698.44/702$			\\
3		& S + R					& \nodata		& \nodata				&& $6.04 \pm 0.01$	& $69.8 \pm 2.7$			&& $6.48 \pm 0.02$	& $7.6^{+0.7}_{-0.6}$			&& $11.3 \pm 0.3$	& $7.6 \pm 0.3$			& $0.40 \pm 0.04$			& $1068.63/704$			\\
4		& S + R					& $6.03^{+0.02}_{-0.03}$& $8.8^{+0.3}_{-0.4}$			&& \nodata		& \nodata				&& $6.35 \pm 0.01$	& $12.0^{+0.8}_{-0.6}$			&& $11.6^{+0.3}_{-0.2}$	& $8.6 \pm 0.3$			& $< 0.08$\tablenotemark{f}		& $870.51/704$			\\
5		& S					& 5.98\tablenotemark{e}	& $1.4^{+7.2}_{-1.4}$			&& $6.09^{+0.02}_{-0.03}$& $41.7^{+3.2}_{-4.2}$			&& $6.49^{+0.01}_{-0.02}$& $6.8^{+0.6}_{-0.9}$			&& $11.2 \pm 0.3$	& $7.6 \pm 0.3$			& 0.28\tablenotemark{e}			& $672.79/690$			\\
6		& S (on)				& \nodata		& \nodata				&& $6.09^{+0.03}_{-0.04}$& $41^{+18}_{-8}$			&& $6.52 \pm 0.02$	& $7.0^{+0.6}_{-0.7}$			&& $10.9 \pm 0.3$	& \nodata			& 0.28\tablenotemark{e}			& $345.72/374$			\\
7		& S (off)				& \nodata		& \nodata				&& $6.08 \pm 0.03$	& $44.2^{+6.5}_{-6.2}$			&& $6.47 \pm 0.03$	& $6.6^{+1.0}_{-0.8}$			&& \nodata		& $8.0^{+0.4}_{-0.3}$		& 0.28\tablenotemark{e}			& $317.14/313$			\\
8		& S\tablenotemark{g}			& 5.98\tablenotemark{e}	& $0.3^{+5.8}_{-0.3}$			&& $6.06^{+0.03}_{-0.02}$& $47.8^{+4.5}_{-9.0}$			&& $6.44^{+0.04}_{-0.03}$& $7.1^{+1.0}_{-0.9}$			&& $11.0 \pm 0.3$	& $7.6 \pm 0.3$			& 0.28\tablenotemark{e}			& $655.01/687$			\\
9		& X + R\tablenotemark{h} 		& $6.06^{+0.02}_{-0.04}$& 18					&& $5.93^{+0.04}_{-0.03}$& $170$				&& $6.43 \pm 0.02$	& 11					&& 10.5\tablenotemark{e}& 10.5\tablenotemark{e}		& \nodata				& $435.86/439$			\\
10		& X + R\tablenotemark{i} 		& $6.30 \pm 0.01$	& $12.5^{+0.6}_{-0.8}$			&& $5.73^{+0.14}_{-0.12}$& $160^{+300}_{-100}$			&& $6.56^{+0.06}_{-0.04}$& $3.8^{+2.0}_{-1.1}$			&& 11.2\tablenotemark{e}& 7.7\tablenotemark{e}		& \nodata				& $442.21/439$			\\
\enddata
\tablecomments{Model 2 is our preferred model.}
\tablenotetext{a}{Extragalactic power-law normalization at 1~\kev\ in \pownorm, assuming a photon index of 1.46.\\ }
\tablenotetext{b}{Emission measure $\mbox{E.M.} = \EMint$.\\ }
\tablenotetext{c}{S = \suzaku; R = \rosat; X = \xmm.\\ }
\tablenotetext{d}{Carbon column density of the \texttt{vphabs} model used to model XIS contamination above that included in the CALDB (see \S\ref{subsec:SuzakuResults} for details).\\ }
\tablenotetext{e}{Value frozen during fitting.\\ }
\tablenotetext{f}{$2\sigma$ upper limit.\\ }
\tablenotetext{g}{Fit with variable abundances (see \S\ref{subsec:Abundances} and Table~\ref{tab:Abundances}).\\ }
\tablenotetext{h}{From \citet{henley07}.\\ }
\tablenotetext{i}{Data reanalyzed to match method used for \suzaku\ (see \S\ref{sec:XMMAnalysis} for details).\\ }
\end{deluxetable}
\clearpage
\end{landscape}

\end{document}